\PassOptionsToPackage{prologue,dvipsnames}{xcolor}
\documentclass[sigconf]{acmart}
\makeatletter
\def\@ACM@checkaffil{
    \if@ACM@instpresent\else
    \ClassWarningNoLine{\@classname}{No institution present for an affiliation}%
    \fi
    \if@ACM@citypresent\else
    \ClassWarningNoLine{\@classname}{No city present for an affiliation}%
    \fi
    \if@ACM@countrypresent\else
        \ClassWarningNoLine{\@classname}{No country present for an affiliation}%
    \fi
}
\makeatother

\AtBeginDocument{%
  \providecommand\BibTeX{{%
    \normalfont B\kern-0.5em{\scshape i\kern-0.25em b}\kern-0.8em\TeX}}}

\setcopyright{acmcopyright}
\copyrightyear{2022}
\acmYear{2022}
\acmDOI{XXXXXXX.XXXXXXX}

\acmConference[SIGMOD ’23]{SIGMOD '23: ACM SIGMOD/PODS Conference}{June 18–23, 2023}{Seattle, WA}
\acmPrice{15.00}
\acmISBN{978-1-4503-XXXX-X/18/06}

\usepackage{balance}
\usepackage{enumitem}
\usepackage{booktabs}
\usepackage{caption}
\usepackage{subcaption}
\usepackage{emptypage}
\usepackage{geometry}
\newcommand{\var}{\texttt}
\usepackage{algorithm}
\usepackage{algpseudocode}
\usepackage{multicol}

\usepackage{xcolor}

\usepackage{listings}
\lstdefinelanguage{Ini}
{
    basicstyle=\ttfamily\small,
    columns=fullflexible,
    morecomment=[s][\color{Orchid}\bfseries]{[}{]},
    morecomment=[l]{\#},
    morecomment=[l]{;},
    commentstyle=\color{gray}\ttfamily,
    morekeywords={},
    otherkeywords={=,:},
    keywordstyle=
}

\usepackage{pifont}
\usepackage{pdfpages}
\begin{document}

\setcounter{section}{0}
\renewcommand{\thetable}{R\arabic{table}}
\renewcommand\thefigure{R\arabic{figure}}
\renewcommand\thesection{R\arabic{section}}
\renewcommand{\theHtable}{Revison.\thetable}
\renewcommand{\theHfigure}{Revison.\thefigure}
\renewcommand{\theHsection}{Revison.\thesection}

\restoregeometry


\setcounter{table}{0}
\setcounter{section}{0}
\setcounter{figure}{0}
\renewcommand{\thetable}{\arabic{table}}
\renewcommand\thefigure{\arabic{figure}}
\renewcommand\thesection{\arabic{section}}


\title{A Unified and Efficient Coordinating Framework  for Autonomous DBMS Tuning
}


\settopmatter{authorsperrow=4}

\author{Xinyi Zhang}
\authornote{Xinyi Zhang and Zhuo Chang contribute equally to this paper}
\authornote{School of CS \& Key Laboratory of High Confidence Software Technologies, Peking University}
\authornote{Database and Storage Laboratory, Damo Academy, Alibaba Group}
\affiliation{%
  \institution{Peking University \& Alibaba Group}
}
\email{zhang\_xinyi@pku.edu.cn}

\author{Zhuo Chang}
\authornotemark[1]
\authornotemark[2]
\authornotemark[3]
\affiliation{%
  \institution{Peking University \& Alibaba Group}
}
\email{z.chang@pku.edu.cn}

\author{Hong Wu}
\authornotemark[3]

\affiliation{%
  \institution{Alibaba Group }
  \streetaddress{Wangjing Tower A, Ali Center, Chaoyang District} 
  \postcode{100102} 
}
\email{hong.wu@alibaba-inc.com}

\author{Yang Li}
\authornotemark[2]
\affiliation{%
  \institution{Peking University}
}
\email{liyang.cs@pku.edu.cn}

\author{Jia Chen}
\authornotemark[2]
\affiliation{%
  \institution{Peking University}
}
\email{MaCasK@stu.pku.edu.cn}

\author{Jian Tan}
\authornotemark[3]
\affiliation{%
  \institution{Alibaba Group}
  \streetaddress{Wangjing Tower A, Ali Center, Chaoyang District} 
  \postcode{100102} 
}
\email{j.tan@alibaba-inc.com}

\author{Feifei Li}
\authornotemark[3]
\affiliation{%
  \institution{Alibaba Group }
  \streetaddress{Wangjing Tower A, Ali Center, Chaoyang District} 
  \postcode{100102} 
  }
\email{lifeifei@alibaba-inc.com}

\author{Bin Cui}
\authornotemark[2]
\authornote{Institute of Computational Social Science, Peking University(Qingdao), China}
\affiliation{%
  \institution{Peking University}
  \postcode{100080}}
\email{bin.cui@pku.edu.cn}

\renewcommand{\shortauthors}{Xinyi Zhang and Zhuo Chang, et al.}

\begin{abstract}
Recently using machine learning (ML) based techniques to optimize the performance of modern database management systems (DBMSs) has attracted intensive interest from both industry and academia.
With an objective to tune a specific component of a DBMS (e.g., index selection, knobs tuning), the ML-based tuning agents have shown to be able to find better configurations than experienced database administrators (DBAs).
However, one critical yet challenging question remains unexplored --
how to make those ML-based tuning agents work collaboratively.
Existing methods do not consider the dependencies among the multiple agents, and the model used by each agent only studies the effect of changing the configurations in a single component.
To tune different components for DBMS, a coordinating mechanism is needed to make the multiple agents be cognizant of each other.
Also, we need to decide how to allocate the limited tuning budget (e.g., time and resources) among the agents to maximize the performance. 
Such a decision is difficult to make since the distribution of the reward (i.e., performance improvement) corresponding to each agent is unknown and non-stationary.
In this paper, we study the above question and present a unified coordinating framework to efficiently utilize existing ML-based agents.
First, we propose a message propagation protocol that specifies the collaboration behaviors for agents and encapsulates the global tuning messages in each agent's model. 
Second, we combine Thompson Sampling, a well-studied reinforcement learning algorithm with a memory buffer so that our framework can allocate the tuning budget judiciously in a non-stationary environment.
Our framework defines the interfaces adapted to a broad class of ML-based tuning agents, yet simple enough for integration with existing implementations and future extensions.
Based on extensive evaluations, we show that this framework can effectively utilize different ML-based agents and 
find better configurations with $1.4$\textasciitilde$14.1\times$ speedups on the workload execution time compared with baselines. 
\end{abstract}



\maketitle

\begin{figure*}
\centering

    \includegraphics{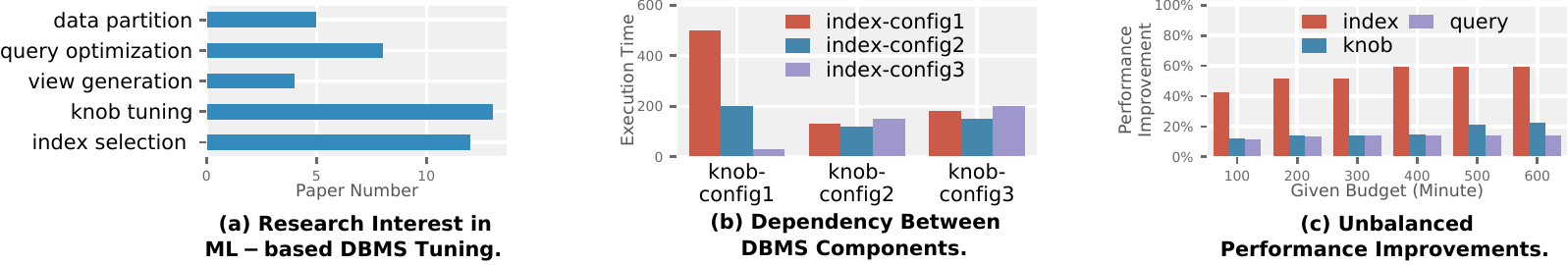}

\caption{Motivating Examples. Figure 1a shows the number of published papers on ML-based tuning agents from 2017 to 2022. Figures 1b presents the query execution times under different configuration settings for knob and index components.
Figure 1c presents the performance improvement of three tuning agents given different tuning budgets on TPC-H workload.}

\label{fig:intro}
\end{figure*}

\section{Introduction}\label{sec:intro}


Database management systems (DBMSs) are playing critical roles in a broad spectrum of data-intensive applications. 
The performance of a DBMS depends on multiple configurable components, e.g., configuration knobs, index settings, and materialized views.
Tuning the DBMS is crucial to obtain high performance, but the process is not trivial.
As a common practice, database administrators (DBAs) put considerable effort into finding appropriate settings for the DBMS.
However, since the NP-hard nature~\cite{DBLP:journals/pvldb/0001Z021} of the tuning problem, manual manipulation can not handle such complex tuning tasks easily.
Due to the shift to cloud environments, there are large-scale instance deployments with increasingly diverse workloads in cloud databases. The manual tuning fails to scale well.

Fortunately, machine learning (ML) techniques could help to overcome this challenge since they can model complex functions and automate the performance tuning process.
Researchers have been extensively working on approaches that use ML techniques to optimize database systems (i.e., ML-based tuning agents). 
Figure \ref{fig:intro}a presents the number of published research focusing on  the ML-based tuning agents for DBMS in recent five years.
These works include index selection~\cite{DBLP:conf/sigmod/DingDM0CN19, DBLP:journals/apin/LicksCMPRM20, DBLP:conf/icde/SadriGL20,DBLP:conf/cikm/LanBP20, DBLP:conf/icde/PereraORB21, DBLP:conf/ideas/SharmaDF21,DBLP:conf/icde/ZhouLLJLWF22, DBLP:conf/edbt/KossmannKS22,DBLP:conf/sigmod/00010SWNCB22,DBLP:journals/dase/WuLZZC22,DBLP:conf/sac/SharmaD22, Gao2022AutomaticIS,DBLP:journals/dase/WuLZZC22,DBLP:journals/jcst/GeCC21}, knobs tuning~\cite{DBLP:conf/sigmod/AkenPGZ17,DBLP:journals/pvldb/TanZLCZZQSCZ19,DBLP:conf/sigmod/ZhangLZLXCXWCLR19,DBLP:journals/pvldb/LiZLG19,DBLP:conf/kdd/FekryCPRH20,DBLP:conf/lifetech/IshiharaS20,DBLP:conf/hotstorage/KanellisAV20,DBLP:conf/sigmod/ZhangWCJT0Z021,DBLP:conf/edbt/GurYSR21,DBLP:journals/pvldb/CeredaVCD21,DBLP:conf/sigmod/ZhangW0T0022, DBLP:journals/pvldb/KanellisDKMCV22, DBLP:conf/sigmod/CaiLZZZLLCYX22},  view generation~\cite{DBLP:conf/icde/Yuan0FSH20,DBLP:conf/icde/Han00S21,DBLP:journals/cn/XiaZRW22,DBLP:journals/corr/abs-1903-01363}, query optimization~\cite{DBLP:journals/pvldb/ZhouLCF21, DBLP:conf/sigmod/WangZYDHDT0022,DBLP:journals/corr/abs-1808-03196,DBLP:conf/sigmod/MarcusP18,DBLP:journals/corr/abs-1808-03196,DBLP:conf/icde/Yu0C020,DBLP:journals/corr/abs-1911-11689,DBLP:conf/kdd/0008Y000CZ022}, and data partition~\cite{DBLP:conf/sigmod/HilprechtBR20, DBLP:conf/sigmod/EldeebCCY22, DBLP:conf/sigmod/DurandPPMBSSRB18,DBLP:conf/adbis/DurandPPBGS19,DBLP:journals/pvldb/ZouDBIYJJ21,DBLP:journals/jcst/HanLL22}.
Given a workload, the ML-based tuning agents aim to improve the database's performance based on certain objective functions (e.g., higher throughput and lower latency).
For example, to tune the configuration knobs in a DBMS, a knobs tuning agent utilizes a ML model to suggest a promising configuration and applies it to the DBMS.
Then the agent updates its model based on the performance of the configuration to  learn the underlying objective function and to improve its tuning policy.

Although the ML-based tuning agents have demonstrated superior performance and efficiency compared to manual tuning~\cite{DBLP:journals/chinaf/HuangQZTLC23,DBLP:journals/dase/LanBP21}, they always focus on the tuning for a single configurable component and  neglect the dependencies between agents.
However, the performance of a DBMS is influenced by multiple components.
Optimal configuration choices for one component would depend on the configurations of the other~\cite{DBLP:journals/debu/PavloBJMMALS19}.
For example, it is better to set a larger query cache, smaller buffer size when no indexes are built, and smaller query cache and larger buffer pool when suitable indexes are built~\cite{DBLP:journals/debu/PavloBJMMALS19}.
Thus, running a standalone agent is very likely to recommend sub-optimal configurations.
Figure \ref{fig:intro}b shows a concrete example when running Q19 from TPC-H workload. 
We observe that the optimal configuration is the combination of knob configuration 1 and index configuration 3.
However, when knob configuration 2 or 3 is applied to DBMS, the index agent would suggest sub-optimal configurations (i.e., index configuration 2).
And when running a tuning agent, the configurations in other components usually are not the ideal ones (e.g., knob configuration 1 in the example).
Hence, tuning multiple components together would own more advantages compared with the standalone tuning.

The multi-component tuning is essentially an optimization problem that navigates a joint configuration space composed of the subspace for individual components.
The composite configuration space is huge and complex, which leads to scalability issues related to the curse of dimensionality ~\cite{DBLP:journals/pvldb/LiSZJLDZY00021}.
Besides, since the configuration spaces for different components are distinct (e.g., binary for index selection, continuous for knob tuning, and tree-structured for query rewrite), it is impossible to jointly optimize them directly.
Meanwhile, space decomposition has shown promising performance when solving a high-dimensional optimization problem~\cite{DBLP:journals/pvldb/LiSZJLDZY00021,DBLP:conf/aaai/0001RV0BSW0G20,DBLP:conf/aaai/LiJGSZ020,DBLP:conf/kdd/LiSJBZZC22}.
Given the above facts, we decompose the joint space according to the corresponding components and resort to the solvers of existing tuning agents to tune the corresponding components.
Consequently, we can take full advantage of the latest research on ML-based tuning agents customized for a specific component.
To this end, we seek to answer a question ---\textit{``how to coordinate existing ML-based tuning agents to configure multiple components in a DBMS?''}


There are three main challenges behind this question.
First, there are emerging ML-based tuning agents with advanced features for different DBMS components.
From a system perspective, without a high-level abstraction of the agents and a general solution, we can not unitize them for multi-component tuning.
\textit{Can we design a unified solution to conveniently support existing ML-based tuning agents and the integration with future extensions (C1)?}
Second, given the dependencies among different components, the tuning agents should be able to adjust their tuning policies according to others' tuning decisions.
However, existing studies tune the DBMS in a standalone way -- the ML models of tuning agents operate under the assumption that the configurations of other components are fixed. 
The agents could not make accurate predictions, if other agents modify their controlled configurations~\cite{DBLP:journals/debu/PavloBJMMALS19}, thus failing to find global promising configurations.
\textit{To leverage existing agents to tune multiple components, we need a mechanism to make the agents communicate and coordinate with each other (C2).}
Third,
running the ML-based agent is expensive, as it needs time and resources to observe the effects of configurations.
However, the distribution of reward  (i.e., performance improvement) corresponding to each agent is unknown.
And, under a given tuning budget (e.g., time), the agents' reward is unbalanced and non-stationary (decaying in most cases, i.e., diminishing marginal utility), as shown in Figures \ref{fig:intro}c.
\textit{Given these facts, we need to explore a judicious policy that can efficiently allocate  budget to the promising tuning agents (C3).}

To address the above challenges, we propose a unified and efficient coordinating framework, UniTune, for orchestrating  the ML-based tuning agents.
We first analyze existing agents for DBMSs and classify them into three categories: Bayesian Optimization (BO) based, Reinforcement Learning (RL) based, and RL-estimator based approaches.
We abstract the logic of different tuning agents and define interfaces to support a wide variety of existing agents and for future extension (addressing \textit{C1}).
Second, to break the standalone tuning limitation, we propose a message propagation protocol that enables collaboration among agents.
It specifies how one agent broadcasts its tuning decision as a message and how other agents receive and process the message.
To update the tuning policy according to the tuning messages, we encapsulate the messages in each agent's model in the form of context features (addressing \textit{C2}).
Third, we formulate the budget allocation problem as a multi-arm bandit problem with each agent referred to an arm and propose a Thompson Sampling based strategy to select promising agents.
This strategy maximizes the tuning performance based on historic observations by judiciously balancing the trade-off between exploitation and exploration.
To avoid the selection misguided by the changing reward in non-stationary environments, we propose a strategy that utilizes a memory buffer to discard the out-of-date observations (addressing \textit{C3}).

To the best of our knowledge, UniTune is the first unified framework using ML techniques to tune multiple components in a DBMS. 
In summary, we make the following contributions.

\begin{itemize}[leftmargin=*]
    \item To tune multiple components in a DBMS efficiently, we discuss how to utilize the existing ML-based tuning agents and design a unified coordinating framework  with high-level abstraction and flexible interfaces.
    \item To enable collaboration among agents, we propose a  message propagation protocol that specifies their message sharing behaviors  and encapsulates tuning messages in agents' models.
    \item  To allocate the limited tuning budget among agents, we propose a Thompson Sampling based strategy to select promising agents based on historical observations.
    It tackles the non-stationary nature of reward via a memory buffer design.
    \item 
    We conduct extensive evaluations on synthetic and real workloads.
    The result shows that UniTune can support different ML-based tuning agents and recommend better configurations with $1.4$\textasciitilde$14.1\times$ speedups on the workload execution time compared with baselines. 
\end{itemize}
The remainder of the paper is organized as follows. 
We discuss related work in Section \ref{sec:related}. We formulate the multi-component tuning problem and review existing ML-based tuning agents in Section \ref{sec:pre}. Then we provide a high-level abstraction for the ML-based tuning agents and give an overview in Section \ref{sec:overview}.
In Section \ref{sec:opt}, we introduce a message propagation protocol to enable the collaboration of tuning agents.
In Section \ref{sec:budget}, we provide a policy to allocate the limited tuning budget among agents.
Finally, we report the evaluation results in Section~\ref{sec:exp} and end this paper with a conclusion.
\section{Related Work}\label{sec:related}

Recently, configuring DBMSs automatically has attracted intensive interest in both industry and academia.
They aim to automate the database tuning tasks and search for better configurations in a data-driven way~\cite{DBLP:journals/chinaf/WangZCYZL22}. 
They adopt Reinforcement Learning (RL) and Bayesian Optimization (BO) algorithms to tune a DBMS in a trial-and-error manner.
Prior works typically focus on a specific tuning task such as index selection, knobs tuning, query rewrite, view generation, and data partition.

\noindent\textbf{Index Selection.}
Indexes on appropriate columns are vital to speed up query execution~\cite{DBLP:journals/pvldb/LuZSPZDHWPL18}.
ML techniques are utilized to select proper indexes from a large number of possible index combinations.
For example, SmartIX~\cite{DBLP:journals/apin/LicksCMPRM20}, MANTIS~\cite{DBLP:conf/ideas/SharmaDF21} and AutoIndex~\cite{DBLP:conf/icde/ZhouLLJLWF22} adopt RL, and DBA-bandits~\cite{DBLP:conf/icde/PereraORB21} adopts $C^2UCB$ algorithm in BO framework.

\noindent\textbf{Knobs Tuning.}
A DBMS has hundreds of configurable knobs, affecting its performance~\cite{DBLP:journals/pvldb/ZhangCLWTLC22}.
To replace the manual tuning that depends on DBA's experience, the DB community has developed lots of ML-based methods to automate this process, such as BO-based~\cite{DBLP:journals/pvldb/DuanTB09, DBLP:conf/sigmod/AkenPGZ17, DBLP:conf/sigmod/KunjirB20, DBLP:conf/kdd/FekryCPRH20, DBLP:conf/sigmod/ZhangWCJT0Z021, DBLP:conf/sigmod/ZhangW0T0022, DBLP:journals/pvldb/CeredaVCD21, DBLP:journals/pvldb/KanellisDKMCV22} and RL-based~\cite{DBLP:conf/sigmod/ZhangLZLXCXWCLR19, DBLP:journals/pvldb/LiZLG19, DBLP:conf/sigmod/CaiLZZZLLCYX22} methods.

\noindent\textbf{Query Rewrite} aims to transform a query into an equivalent one but with higher performance.
It is an NP-hard problem~\cite{DBLP:conf/icde/FinanceG91} with numerous rewrite orders (e.g., different operators and rules).
LearnedRewrite~\cite{DBLP:journals/pvldb/ZhouLCF21} uses a regression model to estimate the benefit of a rewrite node.
And a light-weight RL-based agent, Monte Carlo Tree Search (MCTS) is adopted to interact with the regression model, searching for a better-rewritten query. 

\noindent\textbf{View Generation.}
Materialized views could save redundant computations among queries that share equivalent sub-queries, based on the space-for-time trade-off principle.
Two approaches~\cite{DBLP:conf/icde/Yuan0FSH20, DBLP:conf/icde/Han00S21} use a regression model to estimate the benefit of the different view candidates and queries and suggest view-query pairs for a given workload via a RL-based agent with deep Q-learning algorithm.

\noindent\textbf{Data Partition.}
Partitioning a database can greatly improve the performance of analytical workloads since data-intensive queries can be assigned to multiple machines.
Many approaches ~\cite{DBLP:conf/sigmod/HilprechtBR20,DBLP:conf/sigmod/EldeebCCY22,DBLP:conf/sigmod/DurandPPMBSSRB18,DBLP:conf/adbis/DurandPPBGS19} utilize RL agents to explore different partition keys.
And Li et al.~\cite{DBLP:conf/sigmod/HilprechtBR20} implements a regression model to estimate partition benefits.

We focus on designing a unified coordinating framework of the ML-based tuning agents to tune multiple DBMS components.
The closest work to us is one recent research, UDO~\cite{DBLP:journals/pvldb/WangTB21}.
It proposes to use RL agents to optimize more DBMS components.
It separates the configurations as heavy and light -- heavy parameters have high reconfiguration overheads (e.g., indexes) and light parameters have negligible overheads (e.g., knobs).
It uses a two-layer loop to reduce reconfiguration overheads.
In the outer layer, a RL agent suggests and applies a heavy parameter.
Then, in the inner layer, another RL agent is initialized and iterates for a number of iterations to find a suitable setting for the light parameters.
However, we should allocate the tuning budget according to the expected utility of tuning agents, instead of the reconfiguration overheads.
The two-layer schedule causes a fixed budget allocation pattern: less tuning budget spent on the outer agent.
The fixed schedule leads to a bad overall performance, as shown in our evaluation in Section \ref{sec:exp-end}. 
Different from our general framework, UDO is not designed to coordinate existing tuning agents.
It focuses on customized algorithms to tune the DBMSs.
For example, it proposes an MCTS variant, delayed-HOO to suggest heavy parameters, and a planner to reduce re-configuration overheads by carefully arranging evaluation orders of index configurations.
Those customized algorithms are orthogonal to our work, and they can be considered as another tuning agent and integrated into our framework.

\begin{figure*}   
  \begin{minipage}[b]{0.5\textwidth} 
    \centering   
    \begin{subfigure}{1.0\textwidth}
    \scalebox{0.8}{
    \includegraphics{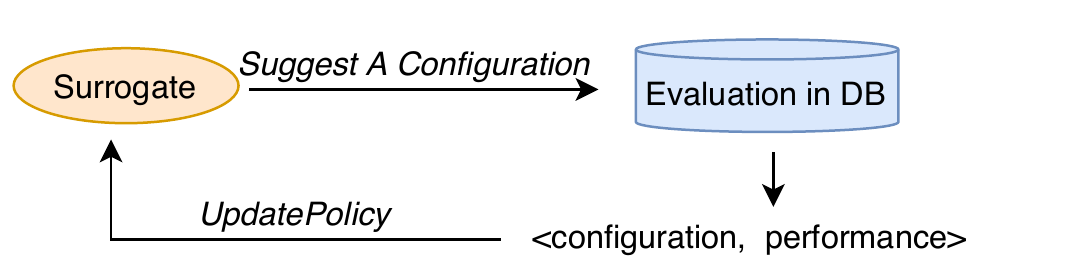}}
     \caption{BO Based Agent.} \label{fig:optimizer1}  
     \end{subfigure}
     \begin{subfigure}{1.0\textwidth}
    \scalebox{0.8}{
    \includegraphics{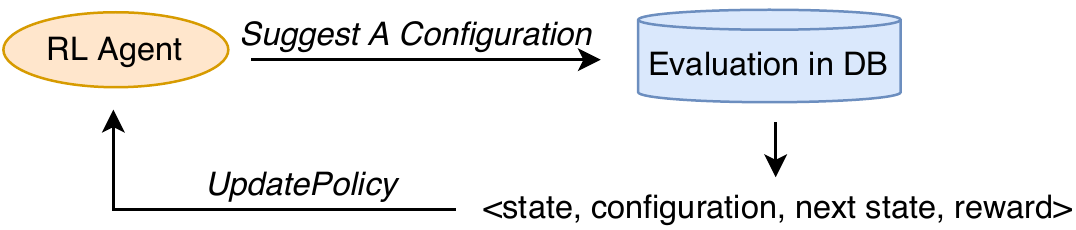}}
    \caption{RL Based Agent.}   
    \label{fig:optimizer2}   
    \end{subfigure}

  \end{minipage}%
  \begin{minipage}[b]{0.5\textwidth}   
    \centering   
    \begin{subfigure}{1.0\textwidth}
    \scalebox{0.8}{
    \includegraphics{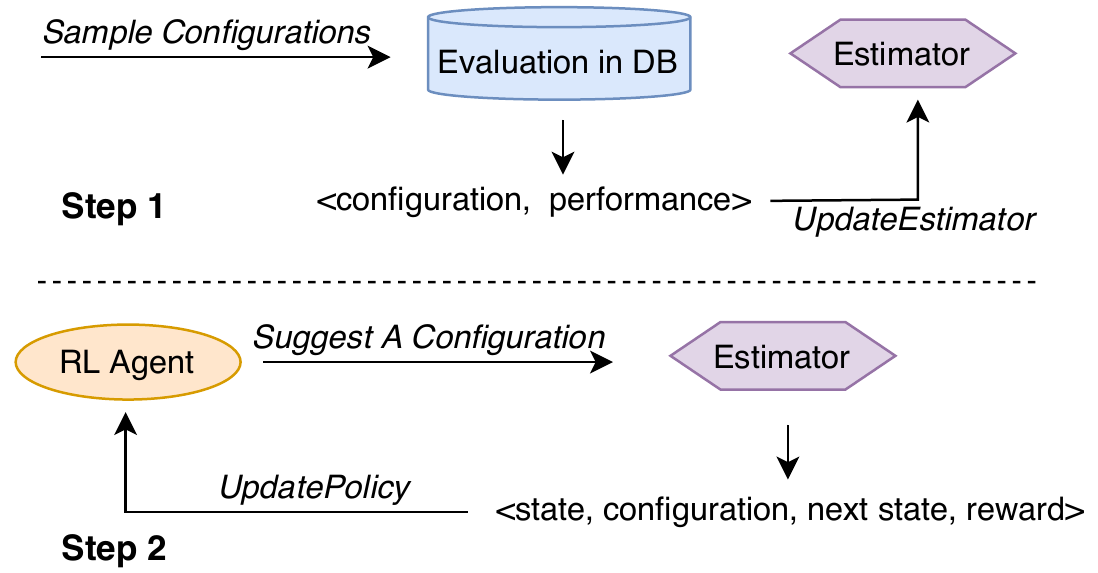} }
    \caption{RL-Estimator Based Agent.}   \label{fig:optimizer3}   
    \end{subfigure}
  
  \end{minipage}   
  \caption{ Three Categories of ML-Based Tuning Agents.} 
  \label{fig:optimizers}
\end{figure*}

\section{Preliminary}\label{sec:pre}
In this section, we formulate the problem and review the existing ML-based tuning agents for DBMS.

\subsection{Terminology and Problem Statement}
We first introduce the associated terminology and then define the multi-component tuning problem.

\noindent\textbf{Tuning Agent.}
An agent is an external system focusing on tuning a specific component of DBMS to improve pre-defined metrics.

\textit{Example 1.} Modern DBMSs have several configurable components that affect their runtime execution and performance, including (1) physical design (e.g., index, view, and data partitioning), (2) configuration knobs, (3) query design (e.g., rewriting a query externally).
An agent is usually responsible for configuring one of these components.
For example, to minimize the execution time of a given workload, a knobs-tuning agent will search for the suitable knobs' values and a view-generation agent will select proper materialized views for queries in the workload.
We denote the agent for component $i$ as $A_i$.


\noindent\textbf{Configuration Subspace.}
A configuration subspace $\Theta_i$ is the set of all possible configurations for a component $i$. 
And we denote the configuration value in  $\Theta_i$ as $\theta_i$.
And the domain of a configuration subspace depends on the nature of the corresponding component and the formulation made by the agent.

\textit{Example 2.} For knob tuning, its domain  could be a mix of continuous variables (e.g., \textit{innodb\_buffer\_pool\_size}), and categorical variables (e.g., \textit{innodb\_stats\_method}). 
For index selection, its domain could be a set of binary values representing whether the indexes are created or not. 
And the domain can even be a set of tuples.
For example, the query rewrite agent makes an atomic tuning decision $(o, r)$, representing applying rewrite rule $r$ on operator $o$.

\noindent\textbf{Configuration Space.}
A configuration space $\Theta$ is a composition of the configuration subspace of $m$ configurable components, i.e., $\Theta = \Theta_1\times  \Theta_2\times ... \times \Theta_m$.

\noindent\textbf{Tuning Budget.}
The tuning budget is the total amount of cost that is allowed to spend on tuning a DBMS.
We can measure the cost from different perspectives, such as time, resource, or money which are positively correlated.
For simplicity, we measure the cost in terms of time in this paper.
We use sub-budget as a hyper-parameter to control the unit of budget spent for executing an agent.

\noindent\textbf{Problem Statement.}
Considering $m$ DBMS components, the goal of multi-component tuning is to optimize a user predefined metric by running  ML-based tuning agents under a limited tuning budget:
$$
 \mathop{\arg\max}_{\theta_1, ..., \theta_m} f(\theta_1, ..., \theta_m; W),
$$
where $\theta_1, ..., \theta_m \in \Theta_1\times  \Theta_2\times ... \times \Theta_m$ is the configuration value suggested by the corresponding agents.
$W$ corresponds to the given workload.
And $f$ is the database performance metrics, which can be any pre-defined objective, e.g., throughput and 99\%th percentile latency.
Note that the result of $f(\cdot)$ can be observed after the evaluations in the DBMS (e.g., stress testing).


\subsection{ Existing ML-based Agents}\label{sec:pre-ana}
We categorize existing ML-based agents into three types: Bayesian Optimization (BO) based, Reinforcement Learning (RL) based, and RL-estimator based.

\noindent\textbf{BO based agents} formulate the tuning task as a black-box optimization problem and update its tuning policy by interacting with the database system, as shown in Figure \ref{fig:optimizer1}. 
To solve the black-box function, the agent works iteratively:
(1) suggest the next configuration to evaluate by computing an acquisition function value that measures the utility of candidate points,
(2) evaluating the suggested configuration by interacting with the DBMS, and (3) updating a probabilistic surrogate model that describes the relationship between configurations and their performance.

\noindent\textbf{RL based agents} formulate the tuning task as a Markov Decision Process (MDP) and update its tuning policy by interacting with the database system, as shown in Figure \ref{fig:optimizer2}.  
An MDP models a problem as a multi-step process. 
At each step, the agent observes the current state $s$ and chooses an action $a\in A(s)$ based on a policy.
The action results in  a reward $r(s,a)$ and a change of state to $s'$.
The RL based agents aim to find a decision-making policy that maximizes cumulative observed reward.

\noindent\textbf{RL-estimator based agents} also formulate the tuning task as an MDP, but the RL agents interact with a pre-trained estimator to update its tuning policy.
The estimator could estimate the utility of a given action. 
Existing work adopts a two-step manner to train estimators and RL agents, as shown in Figure \ref{fig:optimizer3}.
In step 1, the estimator is trained in an offline manner.
And the training data is obtained by randomly sampling configurations and observing their performance by interacting with the DBMS.
In step 2, the RL agent updates the policy by interacting with the estimator instead of evaluating in DBMS.

\underline{\textbf{Scope Illustration.}}
We focus on configuring multiple components in a DBMS via a unified framework for different ML-based tuning agents.
To make our framework clear and general, we clarify some necessary constraints.
(1) We limit the scope of our framework to the support of  the \textit{external} agents, which configure the database components exposed by DBMS’s APIs without having to modify the DBMS's internal implementation.
The external approach is easy to apply to existing DBMSs without software engineering effort to retrofit their architecture. 
(2) At the algorithm level, we focus on offline tuning.
Online tuning can be implemented via a clone and parallelization scheme to stress-test workloads on multiple cloned instances and apply the safe one in the online database~\cite{DBLP:conf/sigmod/CaiLZZZLLCYX22}.
(3)  
Currently, we let users choose one algorithm for tuning one component (we provide default options for each component). For future work, we can support automatic algorithm selection for each component, for instance, deciding which algorithm to use for the knobs component from the candidates such as OtterTune~\cite{DBLP:conf/sigmod/AkenPGZ17}, CDBTune~\cite{DBLP:conf/sigmod/ZhangLZLXCXWCLR19} and CGPTuner~\cite{DBLP:journals/pvldb/CeredaVCD21}.



\section{Overview}\label{sec:overview}
In this section, we highlight the core abstractions in UniTune and discuss the user interface and workflow.

\subsection{Core Abstractions}

Our framework provides a high-level abstraction for the ML-based tuning agents.
The agents share an iterative workflow, following a trial-and-error manner to tune the DBMS.
They utilize a ML model to predict a promising configuration (e.g., building an index), and evaluate the performance of the suggested configuration.
 Based on the evaluation result, they update the model to improve the efficacy for future decision making.
 Following the above paradigm,  our framework defines an abstraction for the ML-based tuning agents.
Concretely, we categorize the tuning agent based on the adopted algorithms and how  they evaluate the suggested configuration, as discussed in Section \ref{sec:pre-ana}.
And, to support existing ML-based tuning agents, we define the following three base classes ({\fontfamily{txtt}\selectfont BO}, {\fontfamily{txtt}\selectfont RL}, and {\fontfamily{txtt}\selectfont RLEstimator}), and provide the corresponding interfaces, as shown in Table \ref{tab:optimizers}.
We summarize the interfaces as follows. 
\begin{itemize}[leftmargin=*]
\item  {\fontfamily{txtt}\selectfont InitModel:} initializes the model used by the tuning agent.
\item  {\fontfamily{txtt}\selectfont Suggest:} infers the defined model to predict a promising configuration.
\item {\fontfamily{txtt}\selectfont UpdatePolicy}: updates the defined model using the input augmented observation.
\item  {\fontfamily{txtt}\selectfont UpdateEstimator:} updates the estimator using the input augmented observation.
\end{itemize}

This abstraction offers flexible support for different tuning algorithms, as it removes the heterogeneity issue in existing ML-based tuning agents.
 For each main class, the tuning agents differ in  how to implement these interfaces of their logic (e.g., which kind of neural networks to adopt in {\fontfamily{txtt}\selectfont InitModel}, and how to predict a promising configuration based on their model in {\fontfamily{txtt}\selectfont Suggest}).
Using these interfaces, it is easy to integrate an agent with existing implementation in UniTune for multi-component tuning.
A user just needs to inherit the corresponding base class and overrides its functions.
For example, to add a  RL tuning agent -- 
 CDBTune~\cite{DBLP:conf/sigmod/ZhangLZLXCXWCLR19}, the user defines an actor-critic network for the agent in {\fontfamily{txtt}\selectfont InitModel}, and implement how to suggest configurations by inferring the actor network in {\fontfamily{txtt}\selectfont Suggest} as well as how to update the model based on the input observation in {\fontfamily{txtt}\selectfont PolicyUpdate}.

Given the user-implemented interfaces, our framework coordinates the agents of different components automatically, hiding
underlying details for users. First, for one agent, our framework
automates its pipelined execution using these functions.
For {\fontfamily{txtt}\selectfont BO} and {\fontfamily{txtt}\selectfont RL}, it evaluates the configuration  output by  {\fontfamily{txtt}\selectfont Suggest} by interacting with the DBMS and inputs the resulting observation to {\fontfamily{txtt}\selectfont PolicyUpdate}.
For {\fontfamily{txtt}\selectfont RLEstimator}, it samples a configuration, evaluates it in the DBMS, and inputs the resulting observation to {\fontfamily{txtt}\selectfont UpdateEstimator}.
Second, our framework schedules the agents to tune multiple components in a DBMS wisely by maintaining the running records of each agent.
It decides how to allocate limited tuning budget among agents automatically (discussed in Section \ref{sec:budget}) and manages the agents following a message propagation protocol to enable collaboration among agents (discussed in Section \ref{sec:opt}).

\begin{table}[]
\caption{\label{tab:optimizers}Abstraction for ML-based tuning agents.}
\scalebox{0.9}{
\begin{tabular}{|c|l|l|}
\hline
Type         & \multicolumn{1}{c|}{Algorithm}                                    & \multicolumn{1}{c|}{Interfaces}                           \\ \hline
{\fontfamily{txtt}\selectfont BO}           & \begin{tabular}[c]{@{}l@{}}Bayesian \\ Optimization\end{tabular}  & \begin{tabular}[c]{@{}l@{}}{\fontfamily{txtt}\selectfont Suggest}, {\fontfamily{txtt}\selectfont ModleInit},\\ {\fontfamily{txtt}\selectfont UpdatePolicy}\end{tabular} \\ \hline
{\fontfamily{txtt}\selectfont RL}         & \begin{tabular}[c]{@{}l@{}}Reinforcement \\ Learning\end{tabular} & \begin{tabular}[c]{@{}l@{}}{\fontfamily{txtt}\selectfont Suggest}, {\fontfamily{txtt}\selectfont ModleInit},\\ {\fontfamily{txtt}\selectfont UpdatePolicy}\end{tabular} \\ \hline
{\fontfamily{txtt}\selectfont RLEstimator} & \begin{tabular}[c]{@{}l@{}}Reinforcement\\  Learning\end{tabular} & \begin{tabular}[c]{@{}l@{}}{\fontfamily{txtt}\selectfont Suggest}, {\fontfamily{txtt}\selectfont ModleInit},\\ {\fontfamily{txtt}\selectfont UpdatePolicy},{\fontfamily{txtt}\selectfont UpdateEstimator}\end{tabular} \\ \hline
\end{tabular}}

\end{table}


\begin{figure}[t]
    \centering
    \begin{minipage}[t]{0.95\linewidth}
    \tiny
\begin{lstlisting}[language={Ini}]
    [Tuning-Setting]
    components = {'index': 'DBA-Bandit', 'knob':'OtterTune', 
                    'query':'LearnedRewrite'}
    tuning_budget = 108000
    performance_metric = 'execution-time'
    \end{lstlisting}

 \end{minipage}
    \caption{An Example of User Interface.}
    \label{fig:user}
\end{figure}

\subsection{User Interface}
To launch a multi-component tuning task, a user only needs to specify the database settings and the tuning setting.
The database setting refers to the connection information and the workload information.
The tuning setting describes the types of DBMS components to be tuned, the corresponding agent, the performance metric, and the overall tuning budget.
For ease of usage, we adopt configuration files to define a task.
The following code in Figure \ref{fig:user} gives an example and we omit the database setting for space constraints.
It defines  a task that  configures index, knobs, and query components via the ML-based tuning agents -- DBA Bandit~\cite{DBLP:conf/icde/PereraORB21}, OtterTune~\cite{DBLP:conf/sigmod/AkenPGZ17} and LearnedRewrite~\cite{DBLP:journals/pvldb/ZhouLCF21}, respectively.
The tuning budget is set to 30 hours, and the tuning objective is the execution time.

\subsection{Workflow}

\begin{figure}
\flushleft 
\scalebox{0.69}{
    \includegraphics{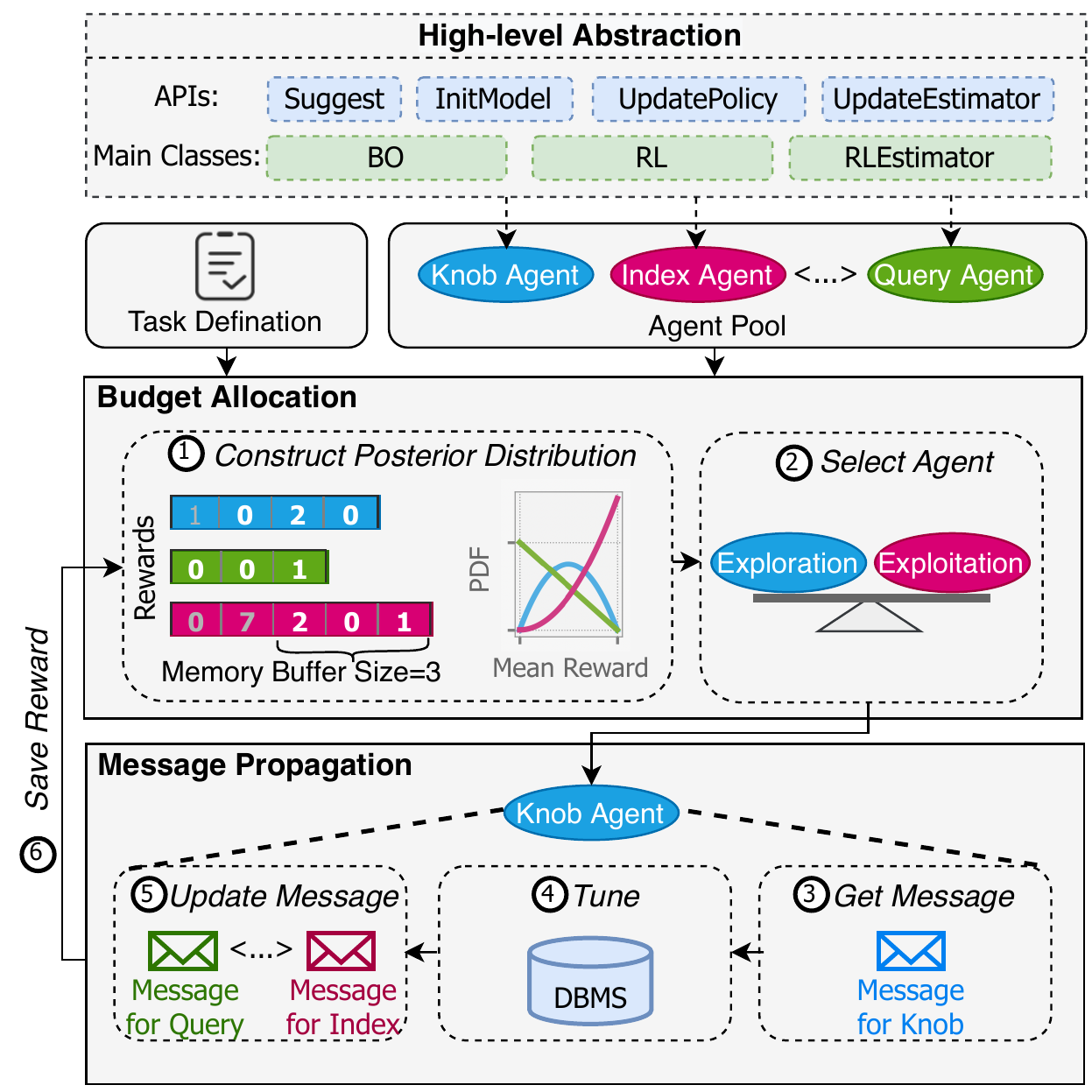}}
\caption{Workflow of UniTune. UniTune has a ``high-level abstraction'' which supports different ML-based tuning agents, a ``budget allocation'' module which selects promising tuning agents, and a ``message propagation'' protocol which enables the collaboration among agents to resolve their dependency.
}
\label{fig:workflow}
\end{figure}

Figure \ref{fig:workflow} presents the overview of UniTune. 
The critical components are emphasized to illustrate its internal workflow.
Given the  tuning agents and the task information defined by a user, the two modules (\textit{budget allocation} and \textit{message propagation}) execute iteratively.
First, the \textit{budget allocation} module selects a promising tuning agent and the selected agent is passed to the \textit{message propagation} module.
Then, the \textit{message propagation}  module arranges the input agent to tune the DBMS while cooperating with the other agents, and finally, the tuning reward obtained is returned to the \textit{budget allocation} module to update the resource allocation strategy.
The workflow ends once the tuning budget is spent out or a termination criterion is met.
We explain the two modules as follows.

\noindent\textbf{Budget Allocation}  selects a promising agent for tuning the DBMS at each step.
The budget allocation problem can be formulated as a multi-arm bandit problem with non-stationary reward (discussed in Section \ref{sec:budget-view}).
The goal is to maximize the cumulative expected reward (i.e., the improvement of a performance metric) within the given budget.
The challenge is that the reward of an agent can only be observed after evaluating its actions using a certain budget, and the reward is non-stationary -- it changes as the optimization proceeds.
To make good decisions, this module judiciously balances exploration and exploitation via Thompson Sampling, a classic algorithm for online decision problems (discussed in Section \ref{sec:budget-win}).
Concretely, it constructs the posterior distributions of reward for each agent based on the reward observations (step 1) and selects an agent based on the probabilities sampled from the posterior distribution (step 2).
To adapt to the non-stationary reward, we use a memory buffer to constrain the number of considered reward observations when constructing the posterior distributions.

\noindent\textbf{Message Propagation} enables collaboration across tuning agents.
The selected agent executes the procedure following a message propagation protocol, which specifies how it should process the received message and broadcast its tuning messages to other agents (discussed in Section ~\ref{sec:opt}).
The agent first obtains its message which contains the  tuning decisions made by other agents (step 3).
Then, the agent encapsulates the message in its model (discussed in Section \ref{sec:opt-context2}) and tunes its responsible component using a sub-budget (step 4).
If it finds a better configuration, it applies the configuration to the DBMS and expresses its tuning decisions in the form of a context feature (discussed in Section \ref{sec:opt-context1}), which is propagated to other agents (step 5).
In the end, we append the observed reward to the corresponding record for later budget allocation (step 6).

Algorithm \ref{alg:top} gives the pseudo-code showing the main optimization loop of our framework.
The sub-budget is a hyper-parameter that controls the amount of budget for running one agent.
It should be larger than the maximal time for the evaluation in DBMS (e.g., stress testing) but as small enough to enable fine-grained budget allocation.
In our implementations, we set it twice the timeout of stress testing time.
UniTune first initializes the current best performance  as the default performance (Line \ref{line-init}).
Then it selects an agent (Line \ref{line-selectopt}).
The selected agent receives the message (i.e., a context feature) sent by other agents (Line \ref{line-getcontext}), configures the DBMS using a sub-budget and returns the best configuration found during tuning and its performance (Line \ref{line-optrun}).
UniTune saves the reward (i.e., performance improvement) (Line \ref{line-save}).
If the selected agent finds a better configuration, it  applies the configuration  and update the message to  other agents (Line \ref{line-better1}-\ref{line-better2}).

\begin{algorithm}[t]
\caption{Top-level Design.}\label{alg:top}
\begin{algorithmic}[1]
\Require Agents $ \{A_i\}^m_{i=1}$, tuning budget $K$, sub-budget $k$.
\Ensure A suggested configuration for best performance.

\Function{Framework}{$ \{A_i\}^m_{i=1}$, $K$}
\State // Initialize global performance as the default one
\State $perf_{glocal} \gets$ init(). \label{line-init}
\While{tuning budget is available}
\State $A_j \gets$ SelectAgent($ \{A_i\}^m_{i=1}$) \label{line-selectopt}.
\State $context\gets$ GetMessage($A_j$). \label{line-getcontext}
\State // Run the selected agent using a sub-budget
\State $config_{inc}, f_{inc} \gets A_j.Run(context, k)$\label{line-optrun}.
\State // Save the performance improvement
\State $A_j.history.append(max(0, f_{inc} - f_{global} ))$\label{line-save}.
\If{$ f_{inc} \leq \var{f}_{\var{global}} $ } \label{line-better1}
\State $ f_{global} \gets f_{inc} $.
\State $A_j.apply(config_{inc})$.
\State // Update messages to other agents
\State UpdateMessage($ \{A_i\}^m_{i=1}-A_j$). \label{line-update}
\EndIf \label{line-better2}
\EndWhile
\EndFunction
\end{algorithmic}
\end{algorithm}

\section{ Message Propagation Protocol}\label{sec:opt}
 To enable collaboration across tuning agents, we propose a message propagation protocol to specify their behaviors.
Within this protocol, the tuning agents execute alternatingly and broadcast their tuning decisions (i.e., messages) to other agents.
In the following, we discuss our design options in Section \ref{sec:opt-design} and introduce
how to represent messages and how to make the agents learn from the messages in Section \ref{sec:opt-design}.
To support RL-estimator based agents with the protocol, we provide a new training paradigm in Section \ref{sec:opt-rl}.

\subsection{Design Options}\label{sec:opt-design}
The ML-based tuning agents use the evaluation in DBMS to obtain the performance of a configuration, as we analyzed.
Therefore, to ensure the validation of the observation, the change of configurations (including the configurations of other components) can not occur during the evaluation so that we can not run two agents simultaneously.
To this end, there are two options for tuning multiple components in a database system.
(1) \textit{joint optimization} that joins the subspaces of DBMS components and runs a single agent to optimize over the joint space, and
(2) \textit{alternating optimization} that runs multiple tuning agents alternatingly to optimize over the corresponding subspaces.

Our protocol adopts alternating optimization.
The joint optimization has severe scalability issues and it relies on a single tuning agent which is hard to adapt to different DBMS components.
First, the number of evaluations needed to reach the global optimum increases exponentially as the dimension of the search space grows. 
The presence of some regions with large posterior uncertainties can result in over-exploration and failure to exploit promising areas. 
Standard Bayesian optimization may perform worse than random search in some high-dimensional spaces~\cite{DBLP:conf/ijcai/WangZHMF13}. 
And reinforcement learning suffers from sparse rewards, and may not learn a successful policy~\cite{DBLP:conf/aaai/WarnellWLS18}.
Second, the underlying objective function is complex, and thus fitting one global model over a huge space is difficult. 
For example, SmartI~\cite{DBLP:journals/apin/LicksCMPRM20}, a RL based agent for index selection, adopts Deep Q-Learning for discrete subspace, and CDBTune~\cite{DBLP:conf/sigmod/ZhangLZLXCXWCLR19}, a RL based agent for knob tuning, adopts DDPG for continuous subspace.
It is non-trivial to jointly optimize in two subspaces by a single agent.
In contrast, alternating optimization solves the optimization problem efficiently by decomposing the full space to subspaces, which is to only configure the selected components through the corresponding tuning agents.
Thus, it can take full advantage of the emerging techniques for tuning a specific component.
For example, we could run two agents alternatingly. 
First, we can apply, e.g., SmartI, by consuming a fraction of the budget, and fix the best indexes found so far to the DBMS. 
Second, we switch to, e.g., CDBTune, and use another fraction of the budget to figure out a best knob that the new agent could find out. 
In UniTune, the budget allocation module decides which agent to execute step by step, as discussed in Section \ref{sec:budget}.
In each step, the alternating optimization essentially fixes some dimensions of the joint space and optimizes over a subspace, which is much smaller than the full space and can be optimized effectively. 
The process searches for the optimum by exploring  different subspaces alternatingly.


\subsection{Message Broadcast via Context}\label{sec:opt-context}
Existing tuning agents work under the assumption that the configurations of other components are fixed.
In UniTune, if an agent finds a better configuration, it applies the configuration to its controlled component and fix the configuration when other agents tune the DBMS. 
However, if one agent modifies its component, the modeling of the other agent will be inaccurate due to environmental changes.
Therefore, we need to broadcast its tuning decisions to other agents, and thus the other agents can be aware of the environmental changes in time.
Then, the next question is how to make the affected agents respond to the environmental changes properly.
One possible response is to initialize their ML models and search from scratch in the new environment.
However, the previous tuning history will be forgotten, and building a new model from scratch needs a large number of observations.
Thus, we turn to another direction -- we make the agents adapt to the environmental changes.
Intuitively, the tuning policies in different environments share certain common knowledge.
We refer to the environmental feature as context.
There exists a mapping from  context and  configuration to the performance metric~\cite{DBLP:conf/sigmod/ZhangW0T0022}.
We expect the same configuration across correlated contexts to have similar performance metrics. 
Utilizing the correlations between contexts can significantly speed up the tuning process.
To this end, we decide to express the messages broadcast among agents in the form of context features and explore two sub-questions: 
(1) how to characterize the context, and 
(2) how to make the agents learn across contexts.

\subsubsection{Context Characterization}\label{sec:opt-context1}
The most direct way is to concatenate all the configurations of other components as a context feature since it indicates the environment changes directly due to other agents' tuning behaviors.
However, the configurations might be unstructured.
For example, the agent for query rewrite applies several (operator, rule) transmissions to a query in the workload.
The number of transmissions is mutative.
Extra feature engineering effort is needed to support the extension of different agents.
Inspired by OtterTune~\cite{DBLP:conf/sigmod/AkenPGZ17}, we resort to the DBMS’s internal runtime metrics to characterize the effect of configurations.
All modern DBMSs expose a large amount of information about the system, such as statistics on the number of pages read/written, query cache utilization, and locking overhead.
And they are affected by the configuration settings.
To observe a context for an agent, we fix the best-ever configurations of other components,  keep the configuration
of its responsible components as default, conduct stress tests on the DBMS to collect the internal metric statistics.

\subsubsection{Context Encapsulation}\label{sec:opt-context2}
To utilize the tuning messages, we need to encapsulate the context feature in the models of tuning agents.
For a BO based agent, we augment the context feature to the input of the surrogate.
Then the surrogate models the mapping from  context and configuration to the performance.
For a RL based agent, we concatenate the context feature to the observed state.
Then the tuning policies, a mapping from state to action, learned by the neural network will be similar among correlated contexts.
For a RL-estimator based agent, we add the context feature to the input of the estimator.
Then the estimator can estimate the utility of given actions in different contexts.
In a summary, the format of the training data for RL agents is $\langle\textrm{state, configuration, next state, reward}\rangle$, where both ``state'' and ``next state'' here contain the concatenated context features.
And the format for  BO and RL-estimator agents is $\langle\textrm{context feature, configuration, performance}\rangle$.


\subsection{Uncertainty-aware Training for RL-estimator Based Agents}\label{sec:opt-rl}
In alternating optimization, if one tuning agent finds a better configuration, we apply it to the DBMS and broadcast its tuning decisions to other agents.
Then, other agents could update their suggestions based on the received message, moving towards a better configuration composition.
It is straightforward for BO based and RL-based agents since they allocate sub-budget to update tuning policies and broadcast the tuning messages when they are invoked.
However, for  RL-estimator based agents, in the existing two-step manner, as shown in Figure \ref{fig:optimizer3}, they first use the allocated sub-budget to train an estimator by the expensive evaluation in DBMS (step 1).
After the estimator is trained, the RL agent updates its tuning policy and only recommends a final configuration (step 2).
Therefore,  during step 1 (most of the time), the other agents lose the opportunity to refine their suggestions according to the tuning decisions of the RL-estimator agent.
In addition, the distribution of randomly sampled configurations in step 1 may not coincide with the configurations inferred by the RL agent in step 2, leading to inaccurate estimation.

To address these issues,  we design an uncertainty-aware training schema for RL-estimator based agents.
We train the RL agent and estimator in one step, enabling the RL agent to suggest promising configurations every time the agent is invoked.
Specifically, we adopt an uncertainty-aware surrogate to estimate the uncertainty of the configurations, such as the Gaussian citation.
When a RL-estimator based agent is invoked, we first train the RL agent, inferring the estimator.
And we put the inferred configuration in a priority queue with estimated uncertainty as the priority.
After the RL agent converges, we evaluate its recommended configuration by interacting with the DBMS.
If a better configuration is found, we apply it and broadcast it to other agents, facilitating cooperation among agents.
Then we use the left sub-budget to evaluate the performance of the configurations in the priority queue and update the estimator based on the augmented observations.
The uncertainty-aware schema guides the sampling of configurations in a demand-oriented way and leads to more accurate prediction from the estimator.

\begin{figure*}
     \centering
     \begin{subfigure}[b]{0.2\textwidth}
         \centering
         \includegraphics{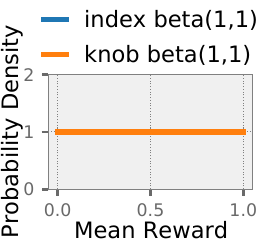}
         \caption{Epoch 0.}
         \label{fig:itero}
     \end{subfigure}
  \hspace{-1em}
     \begin{subfigure}[b]{0.2\textwidth}
         \centering
         \includegraphics{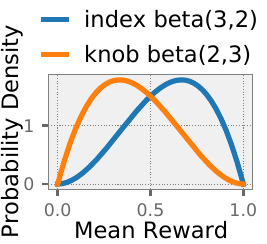}
         \caption{Epoch 4.}
         \label{fig:iter4}
     \end{subfigure}
\hspace{-1em}
     \begin{subfigure}[b]{0.2\textwidth}
         \centering
         \includegraphics{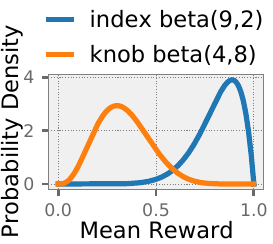}
         \caption{Epoch 10.}
         \label{fig:iter10}
     \end{subfigure}
\hspace{-1em}
     \begin{subfigure}[b]{0.2\textwidth}
         \centering
         \includegraphics{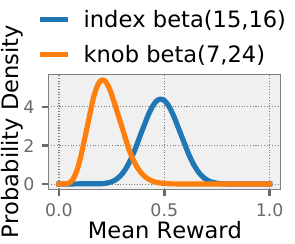}
         \caption{Epoch 30.}
         \label{fig:iter30}
     \end{subfigure}
\hspace{-1em}
     \begin{subfigure}[b]{0.2\textwidth}
         \centering
         \includegraphics{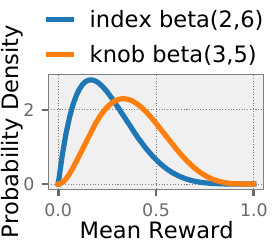}
         \caption{Epoch 30 with  Buffer.}
         \label{fig:iter30-w}
     \end{subfigure}
        \caption{Posterior Distribution in Different Epochs. We present probability density functions over mean reward.}
        \label{fig:ts}
\end{figure*}

\section{Budget Allocation}\label{sec:budget}
Various policies can be employed to allocate a tuning budget among agents.
One simple strategy is a round-robin that equally allocates the tuning budget in turns.
However, the reward of different agents often varies dramatically.
For example, some workloads are very sensitive to the index configuration while query rewrite would offer little or even no improvement.
Therefore, we should spend more budget on learning good indexes instead of rewriting the queries.
The round-robin would waste the budget on the agents which turn out to have little reward. 
To wisely allocate the budget, we find that agent selection can be defined as a multi-armed bandit (MAB) problem~\cite{DBLP:journals/ftml/BubeckC12} from statistical machine learning, where different arms refer to the tuning agents.
In the following, we discuss the formulation of this problem and present a Thompson Sampling based strategy with a memory buffer to solve the MAB.


\subsection{A MAB View for Agent Selection}\label{sec:budget-view}
The MAB problem has been a subject of intense studies in statistics for decades.
The name comes from imagining a gambler at a slot machine with multiple arms, who has to decide which arm to pull~\cite{DBLP:journals/ftml/BubeckC12}.
Each time an arm is pulled, the gambler receives a payout.
Because the distribution of payouts corresponding to each arm is not listed, the gambler can learn it only by experimenting. 
Agent selection problem shares a similar essence with MAB since they both sequentially make online decisions to maximize a total payoff from unknown distributions.

\noindent\textbf{MAB  for Agent Selection.}
We formally define the stochastic MAB problem in the agent selection case.
We are given a set of tuning agents $\mathbf{A}=\{A_i\}_{i=1}^m$ and each agent is an ``arm''.
Let $\mathbf{T}=\{1,2,...,T\}$ denotes a sequence of decision epochs ($T$ can be derived from the setting of tuning budget and sub-budget).
At each decision epoch, we must select an agent to execute.
After selecting an agent $A_j\in\mathbf{A}$ at epoch $t\in\mathbf{T}$, a real-valued reward $X_t^j\in\mathbb{R}$  is observed, where $X_t^j$ is a random variable with expectation $\mu^j_t=\mathbb{E}[X_t^j]$.
The goal is to maximize the expected cumulative reward in epochs $\mathbf{T}$:
$$\mathop{\arg\max}_{j(1),j(2),\ldots,j(T)}\mathbb{E}\left[\sum_{t=1}^T{\mu_{t}^{j(t)}}\right],$$ 
where $j(t)$ is the arm selected in step $t$.

\noindent\textbf{Reward Function.}
Since we aim to find a configuration composition that optimizes the given performance metric, the semantics of the reward has to be consistent with the tuning goal.
Therefore, we define the reward as the improvement of the performance metric when running the agent.
Specifically, the reward is
$$
X^j_t = max\left(0, f^j_{t,inc} - f_{t,global}\right),
$$
where $f^j_{t, inc}$ denotes the best performance achieved by agent $A_j$ at epoch $t$, and $f_{t,global}$ denotes the best performance achieved before  $t$, assuming that we want to maximize the performance metric $f$.

\underline{\textbf{Technical Challenges:}}
To maximize the cumulative reward, we face two challenges.
The first is the common dilemma in the MAB problem -- the trade-off between exploitation and exploration: 
\begin{itemize}[leftmargin=*]
    \item Exploitation:  one may want to allocate budget to the agents that already yielded high reward in the past. 
    \item Exploration: one may also want to allocate budget to the agents that might earn higher reward in the future.
\end{itemize}
The second is a trade-off between ``remembering' and ``forgetting''  caused by the non-stationary reward in the agent selection case.
In the conventional MABs, the reward distributions do not change over time.
However, as we analyzed, the reward distribution of an agent is not stationary -- it changes as the optimization proceeds. 
In general, the reward of an agent decays since the performance improvement achieved by the agent tends to be saturated as the budget is consumed (i.e., the decreasing marginal returns).
Therefore, the old experience might be no longer applicable to estimating future reward.
Then, to select a promising agent, we should carefully balance the trade-off between remembering and forgetting: 
\begin{itemize}[leftmargin=*]
    \item Remembering: one may want to keep track of more observations to decrease the variance of reward estimates. 
    \item Forgetting: one may also want to dismiss ``old'' information which is less relevant due to possible changes in the underlying reward.
\end{itemize}

\subsection{Thompson Sampling with Memory Buffer}\label{sec:budget-win}
To solve the above challenges, we propose a Thompson Sampling based strategy with a memory buffer.
Thompson Sampling~\cite{DBLP:journals/ftml/RussoRKOW18} is a classic reinforcement learning algorithm for the MAB problem.
At a high level, it chooses an arm to play according to its probability of being the best arm.
It builds up experience (i.e., the observation of selected agents and their reward) to construct a posterior distribution of reward.
And, at each epoch, the algorithm selects an arm according to its sampled probability of being the arm with the highest reward.

\begin{algorithm}[t]
\caption{Agent Selection.}\label{alg:select}
\begin{algorithmic}[1]
\Require Agents $ \{A_i\}^m_{i=1}$, buffer size $s$, a scale factor $rfactor$.
\Ensure A selected agent.

\Function{AgentSelection}{$ \{A_i\}^m_{i=1}$}
\For{$i\gets 1,2,\cdots,m$}\label{line-window1}
    \State $S\gets 0, F\gets 0$ \label{line-initbeta}
    \For{$reward \in A_i.history[-s:]$}
    \If{$reward > 0$}
        \State  $S\gets S+ round\left(\frac{reward}{rfactor}\right)$.\label{line-scale}
    \Else
        \State $F\gets F + 1$.\label{line-window2}
    \EndIf
    \EndFor
    \State Sample $w_i$ from the $Beta(S+1, F+1)$ distribution.\label{line-sample}
\EndFor
\State \Return $\mathop{\arg\max}_{i=1}^m w_i $.\label{line-return}
\EndFunction
\end{algorithmic}
\end{algorithm}

\noindent\underline{\textbf{Exploitation \& Exploration}}. Thompson Sampling addresses the first challenge by sampling the arm from the posterior distribution instead of greedily selecting the arm with the highest expected reward.
Consider the example of running two agents -- index agent and knob agent.
Figure \ref{fig:ts} presents their changing posterior distributions across epochs.
These distributions represent the beliefs of the arms given the observed history.
At epoch 0 (Figure \ref{fig:itero}), we do not have any observation for the reward. Therefore, these two agents will be selected with the same probability.
At epoch 4 (Figure \ref{fig:iter4}), the index agent has gained more reward.
And a greedy policy will select the index agent since it has a larger expected reward.
But in Thompson sampling, the knob agent can still be selected, since the selected arm is sampled from the posterior distribution.
At epoch 10 (Figure \ref{fig:iter10}), the index agent is more likely to be selected, since the posterior is more differentiated with smaller uncertainty (but the knob agent still has the possibility to be selected by Thompson sampling).
Intuitively, if we want to maximize exploration, one would choose the agent entirely at random.
If we want to maximize exploitation, one would greedily choose the agent with the highest expected reward.
Sampling from the posterior distributions achieves a balance between the two goals.

\noindent\underline{\textbf{Remembering \& Forgetting.}}
However, recalling the second challenge that the reward distribution is non-stationary in our case, the conventional Thompson Sampling would be less effective.
Continuing with the example in Figure \ref{fig:ts}, we assume that the reward of the index agent is saturated after epoch 10 (i.e., the index agent will consecutively fail to find better configurations).
But given the large reward of the index agent in the first 10 epochs, its expected reward at epoch 30 (Figure \ref{fig:iter30}) is still larger than that of the knob agent with relatively small uncertainty.
Therefore,  the index agent is more likely to be selected although it offers no future reward.
To address the second challenge, we adopt a simple yet effective strategy -- we use a memory buffer to control the number of considered observations for constructing the posterior distribution.
Given the observed reward  $\{X_t^j\}_{t=0}^{T_j}$ for agent $A_j$, we only utilize $\{X_t^j\}_{t=T_j-s+1}^{T_j}$ (i.e., the previous $s$ observations) to construct its posterior distribution, where $s$ is a memory buffer size.
The buffer size represents the number of considered observations.
With a smaller buffer size, the algorithm will have a lower risk of being biased by past observations and  can better adapt to the changes in future reward.
But, at the same time, it may forget useful knowledge,  missing the opportunity for exploitation.
The buffer size controls the trade-off between forgetting and remembering.
And it has been theoretically proved that sub-linear regret is achievable in non-stationary environments with an appropriate trade-off between forgetting and remembering~\cite{DBLP:conf/nips/GurZB14}.
Figure \ref{fig:iter30-w} presents the posterior distribution with a buffer size of seven. 
The expected reward of the knob agent is higher than that of the index agent and will be more likely to be selected since it gains more reward in the considered previous seven epochs.




As we discussed, there are dependencies among different tuning agents.
The allocation strategy explores different dependency directions through different tuning orders of components. 
It decides the tuning orders  step by step intrinsically by selecting a promising agent at each iteration.
For example, UniTune might start by evaluating one tuning order, if the improvement is small, it explores other directions.
Thus, different tuning orders are explored adaptively based on historic feedback.
Algorithm \ref{alg:select} presents a formal description of the budget allocation strategy in UniTune.
We adopt the beta distribution to describe the posterior distribution of reward, which is widely used in Thompson Sampling~\cite{DBLP:journals/jmlr/AgrawalG12}.
Beta distribution has two parameters, $S$ and $F$ to control the estimation of the expected reward.
As shown in Figure \ref{fig:ts},  after pulling an arm, the posterior distribution of the expected reward can be constructed by simply adjusting the two parameters.
And higher the $S$ and $F$ are, the tighter the concentration of $Beta(S, F)$ is around the mean.
To trade off between remembering and forgetting, we utilize the observed reward in a memory buffer with size $s$ to update the two parameters (Line \ref{line-window1}-\ref{line-window2}).
Then, we sample from the posterior distributions (Line \ref{line-sample}) and select an agent according to the probability of its mean being the largest (Line \ref{line-return}).
The conventional Thompson Sampling observes binary reward and hence is not directly applicable to our case of continuous reward.
``Probabilistic reward''~\cite{DBLP:journals/jmlr/AgrawalG12} is used to adapt Thompson Sampling to the continuous reward scenario -- it scales the continuous reward to a domain of $[0,1]$ and samples a binary reward from a Bernoulli distribution with the scaled reward as its success probability.
However, the ``probabilistic reward'' cause the agent with a positive observed reward possibly to have zero sampled reward.
The information loss compromises the already sparse reward in our case.
Based on our empirical experience, we scale the observed reward by a constant factor and use the rounded value to update  $S$ (Line \ref{line-scale}).

\section{Experimental Evaluation}\label{sec:exp}
\begin{figure*}
\centering
\includegraphics{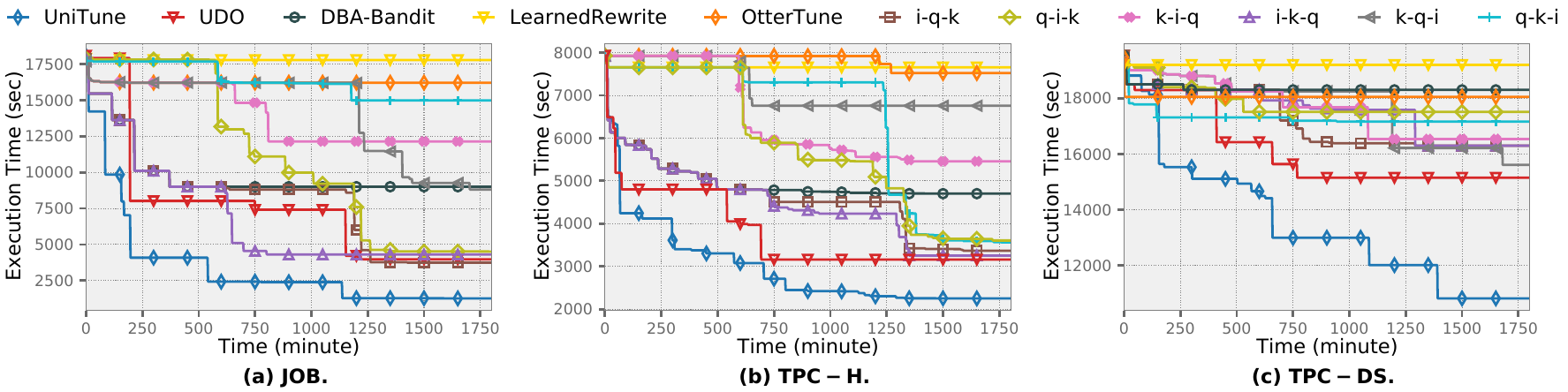}
\caption{The best performance achieved over time by different baselines (bottom left is better).}
\label{fig:exp-end}
\end{figure*}

We conduct experiments to evaluate UniTune.
We describe experimental setup in Section \ref{sec:exp-setup}, conduct end-to-end comparisons with baselines in Section \ref{fig:exp-end}, analyze UniTune in Section \ref{sec:exp-ana}, and present case studies in Section \ref{sec:exp-case}.

\subsection{Experimental Setup}\label{sec:exp-setup}

\noindent\textbf{Agents.}
In most of our experiments, we tune three components in DBMSs -- index, knobs, and SQL query.
Unless stated otherwise, we adopt a BO based agent, OtterTune~\cite{DBLP:conf/sigmod/AkenPGZ17} for knobs tuning, a BO based agent, DBA-Bandit~\cite{DBLP:conf/icde/PereraORB21} for index selection, and a RL-estimator based agent, LearnedRewrite~\cite{DBLP:journals/pvldb/ZhouLCF21} for query rewrite.
To showcase UniTune's extendability, we  replace OtterTune with other knobs tuning agents: CDBTune~\cite{DBLP:conf/sigmod/ZhangLZLXCXWCLR19} and MysqlTuner~\cite{MySQLTuner}.
We also  add a RL-estimator based agent, AutoView~\cite{DBLP:conf/icde/Han00S21} to tune view component.

\noindent\textbf{Workloads.}
We consider three analytic benchmarks: JOB~\cite{DBLP:journals/pvldb/LeisGMBK015},  TPC-H~\cite{tpch} with a scale factor 10 and TPC-DS~\cite{tpcds} with a scale factor 50 since they are widely used in the evaluations for the tuning agents~\cite{DBLP:journals/pvldb/LiZLG19,DBLP:journals/pvldb/KossmannHJS20,DBLP:conf/sigmod/ZhangW0T0022,DBLP:conf/icde/PereraORB21,DBLP:conf/icde/Han00S21}.
And most agents for query rewrite or view generation target at the OLAP scenarios.
The JOB dataset represents a typical real database with unbalanced data distributions. 
We use its provided 113 queries in the workload.
TPC-H  and TPC-DS are both decision support benchmarks that model a real-world data warehousing environment.
For TPC-H, some  queries have costs orders of magnitude higher than the average value.
Adding them in the workload makes the index selection problem much simpler~\cite{DBLP:journals/pvldb/KossmannHJS20} because an index that decreases the cost of at least one of these queries would always outperform indexes for other queries by orders of magnitude.
Therefore, we adopt 12 TPC-H queries, which include a mix of common execution patterns (i.e., pipelines, jobs with multiple joins and filters, and groupby aggregations).
We also add three badly-written queries from the synthetic query set used in LearnedRewrite~\cite{DBLP:journals/pvldb/ZhouLCF21} to  TPC-H workload.
For TPC-DS, we use its provided 99 queries as the workload.

\noindent\textbf{Performance metric and configuration space}.
We optimize the total query execution time of a given workload.
For each workload, we generate an index candidate set as the set of all potentially useful indexes (e.g., columns appearing in the predicates)\cite{DBLP:conf/vldb/ChaudhuriN97} and it serves as a starting point for index selection tool which picks a subset of these indexes. We only consider single-column indexes and generate 59 index candidates for JOB, 54 index candidates for TPC-H, and 237 index candidates for TPC-DS.
We set  a 1500 MB storage budget on built indexes for JOB and TPCH  and  set a 20000 MB budget  for TPC-DS since it has  a larger dataset and much more complex schema.
For the three workloads, we tune 50 knobs.
We utilize the query rewrite rules in Calcite~\cite{DBLP:conf/sigmod/BegoliCHML18}, as LearnedRewrite\cite{DBLP:journals/pvldb/ZhouLCF21} stated.
As for view generation, we consider 29 view candidates for JOB, following the view generation module in Autoview~\cite{DBLP:conf/icde/Han00S21}.

\noindent\textbf{Baselines.} 
The baselines used in the end-to-end evaluation are listed as follows:
\begin{itemize}[leftmargin=*]
    \item \textbf{Standalone Tuning}  tunes a single component with a single agent, as previous studies did. 
    We compare three standalone baselines: DBA-Bandit~\cite{DBLP:conf/icde/PereraORB21}, LearnedRewrite~\cite{DBLP:journals/pvldb/ZhouLCF21} and OtterTune~\cite{DBLP:conf/sigmod/AkenPGZ17}.
    \item \textbf{Sequential Tuning} is a native way to tune multiple components in a DBMS.
    Given a tuning budget, a DBA could allocate it equally to the tuning agents and run them in turn.
    We compare all the permutations for the execution order: i-q-k, i-k-q, q-i-k, q-k-i, k-i-q, and k-q-i. (i-k-q denotes index-knob-query, and so forth.)
    
    \item \textbf{UDO~\cite{DBLP:journals/pvldb/WangTB21}} adopte a  two-layer schema, separating tuning for the heavy and light parameters.
    In the outer layer, an agent for heavy parameters suggests and applies a heavy configuration. 
    Then, in the inner layer, agents for light parameters iterate for a fixed number of inner iterations -- searching suitable light configurations under the applied heavy configuration and evaluating them.
    The best-evaluated performance achieved in the inner loop is considered the performance of the heavy configuration and is used to update the outer agent.
    We adopt the implementation released by the authors ~\cite{udo}.
    For the fairness of comparison, we add LearnedRewrite as the agent for query rewrite, which is not currently supported by UDO. 
    Following the paper~\cite{DBLP:journals/pvldb/WangTB21},  we consider the query agent as the inner agent, since it does not configure the  physical structure of a DBMS.
    The query agent rewrites queries to reduce their execution time before tuning knobs.
    And we set the number for its inner iterations to three based on experiments.
    Therefore, the numbers of inner iterations are three for the query agent and five (the same as the released implementation~\cite{udo}) for the knob agent  on each index configuration.

\end{itemize}

\noindent\textbf{Setting.}
We optimize a MySQL database  deployed on a cloud ECS instance with 16 vCPU and 32GB RAM.
We use the official MySQL default configuration as the initial configuration with no indexes and views built.
We set the tuning budget to 30 hours and set a 10-minute timeout for running one query.
For UniTune, we set the sub-budget to 20 minutes and use a memory buffer with size seven by default.
We also vary the setting of buffer size, to test its robustness.
When allocating the first nine sub-budgets (i.e., initialized phase), UniTune runs the agents in a round-robin manner to bootstrap the Thompson Sampling.
To set $rfactor$ for Algorithm \ref{alg:select}, we observe the maximal reward of executing an agent (i.e., $r_{max}$) for a sub-budget in the initialized phase and set the $rfactor$ as $ \frac{r_{max}}{20}$, resulting in scaled reward approximately ranging  from 0 to 20.

\subsection{End-to-end Evaluation}\label{sec:exp-end}
\begin{figure}
     \centering
      \begin{subfigure}[b]{0.5\textwidth}
         \centering
         \includegraphics[width=0.9\textwidth]{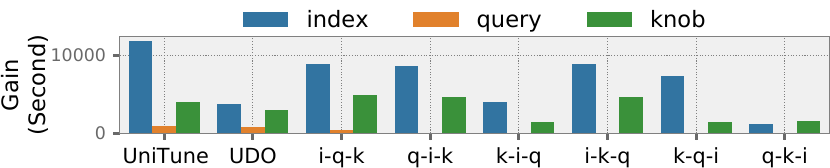}
         \caption{JOB.}
     \end{subfigure}
     \hspace{-1em}
     \begin{subfigure}[b]{0.5\textwidth}
         \centering
         \includegraphics[width=0.9\textwidth]{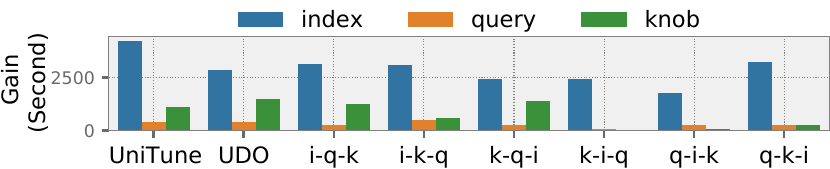}
         \caption{TPC-H.}
     \end{subfigure}
  \hspace{-1em}
    
     \begin{subfigure}[b]{0.5\textwidth}
         \centering
         \includegraphics[width=0.9\textwidth]{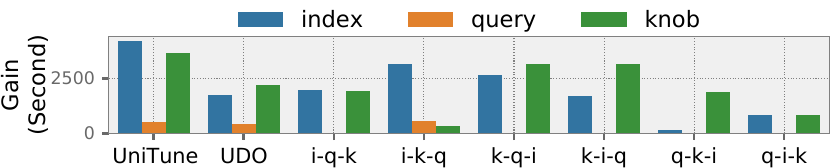}
         \caption{TPC-DS.}
     \end{subfigure}
\caption{Performance breakdown in terms of tuning agents.}
\vspace{-0.5em}
\label{fig:exp-end-ar}
\end{figure}

We compare  UniTune with the baselines.
Figure \ref{fig:exp-end} presents the best performances they achieved over time.
And Figure \ref{fig:exp-end-ar} breaks down the performance improvement of the three agents for multiple component tuning.
We observe that UniTune finds the best configurations on both workloads, achieving $1.4\times$\textasciitilde$14.1\times$ speedups on the execution time of the best configurations compared with baselines.
And no matter how much the tuning budget (x-axis) is, UniTune outperforms the baseline approaches.
The standalone tuning performs worse than the multiple components tuning in general, since they could only find sub-optimal configurations in a subspace. 
The six sequential approaches using different tuning orders converge differently.
As shown in Figure \ref{fig:exp-end-ar}, different orders lead to distinct improvements achieved by the three agents, indicating their dependencies.
And considering the three workloads, there is no clear winner among the orders.
But tuning indexes before knobs generally leads to better performance.

\begin{figure}
\centering
\includegraphics{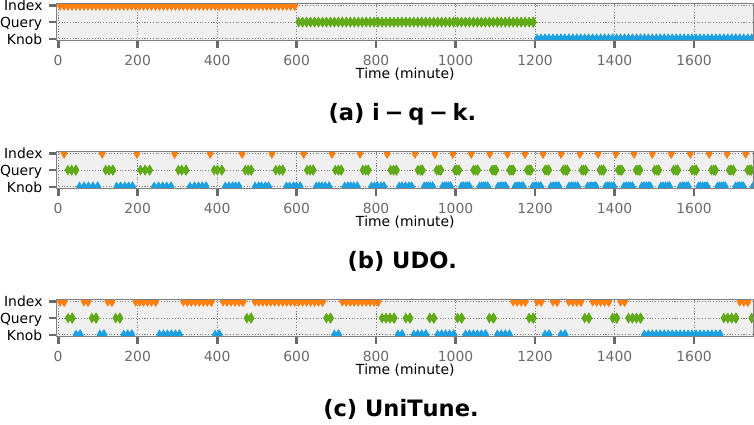}
\caption{Budget allocation patterns of one sequential tuning
approach (i-q-k), UDO, and UniTune on the index, query, and knob agents over time on JOB. (The color block indicates that the budget is allocated to the corresponding agent at the corresponding time slot.)}
\label{fig:exp-end-budget}
\vspace{-0.5em}
\end{figure}

To deep dive, we analyze the baselines tuning multiple components, as shown in Figure \ref{fig:exp-end-budget}.
We observe that UniTune learns a proper allocation strategy by exploiting the historical observations while exploring the less-run agents.
It allocates more budget on index and knob instead of query since tuning indexes and knobs offers more improvement compared with rewriting the queries for the benchmark workload.
And the  number of configurations explored by UniTune  on index, query, knob is approximately 4:2:3.
In the first half period of tuning, UniTune learns to optimize more on the promising components (i.e., index), since it offers more tuning benefit, as shown in Figure \ref{fig:exp-end-ar}.
Then,  it  explores other agents in the second half period as the performance improvement of tuning indexes tends to become saturated.
UDO outperforms sequential approaches in most cases, especially  when the data size is larger (i.e., on TPC-H and TPC-DS).
The index agent in UDO reorders the index configurations and largely reduces the  reconfiguration overheads.
 However, UDO performs worse than UniTune.
 Its  two-layer schema can not flexibly invest  the tuning budget on  promising agents and causes insufficient budget allocated on the outer agent (i.e., index agent), as shown in Figure \ref{fig:exp-end-budget}(b). 
UDO applies one index configuration in the outer layer and executes the query agent for three inner iterations and the knob agent for five inner iterations.
The ratio of configurations explored by UDO on index, query, knob is fixed to 1:3:5 during tuning, mismatching the situation that spending more budget on tuning indexes is beneficial.

\subsection{ Analysis of UniTune}\label{sec:exp-ana}

We carefully design UniTune with message propagation protocol and a budget allocation strategy.
In this section, we analyze UniTune's execution time, evaluate the corresponding designs via
ablation study and variants comparison and then validate the robustness of UniTune on the setting of memory buffer size.

\subsubsection{Execution Time Breakdown.}\label{sec:exp-an-time}
The total execution time of UniTune contains the time for agent selection and the time for agent execution.
The latter is controlled by the sub-budget during which the selected agent tunes the DBMS and updates its tuning policy.
The former is  negligible since it follows closed-form equations in Algorithm \ref{alg:select} without training ML models.
It takes always less than 2 milliseconds in our experiments.

\subsubsection{Ablation Study of Context Features.}
\begin{figure}
\centering
\includegraphics{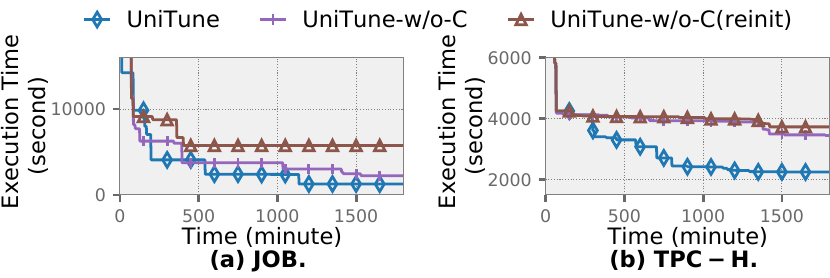}
\caption{Ablation study of context features.}
\label{fig:exp-ana-context}
\vspace{-0.5em}
\end{figure}

We encapsulate context features in the agents' models to learn the tuning policy.
To validate the function of context features, we compare other solutions without context features, including 
(1) UniTune-w/o-C(reinit), which reinitializes an agent's model when its background environment changes (i.e., the other agents change the configuration of their components), as discussed in Section \ref{sec:opt-context}. 
(2) UniTune-w/o-C, which ignores the environmental changes.
As shown in Figure \ref{fig:exp-ana-context}, UniTune outperforms the solutions without context features.
UniTune-w/o-C(reinit) performs the worst since its relearning method  loses previous knowledge. 
Compared with UniTune-w/o-C, UniTune offers more advantages, indicating the relationship between configurations and performance has  variations in different environments.

\subsubsection{Comparison of Different Budget Allocation Strategies.}\label{sec:exp-ana-utility}
\begin{figure}
\centering
\includegraphics{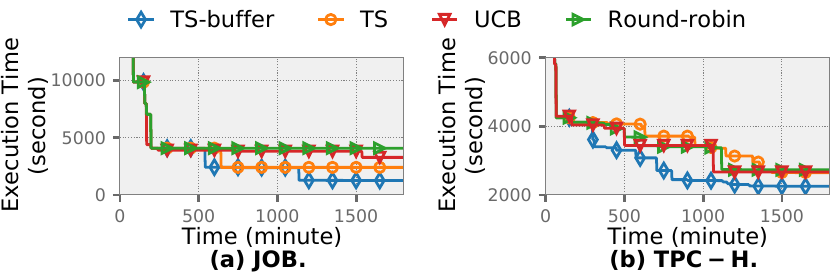}
\caption{Comparison of different budget allocation strategies. TS-buffer is the strategy adopted in UniTune.}
\label{fig:exp-ana-budget1}
\vspace{-0.5em}
\end{figure}

\begin{figure}
\centering
\includegraphics{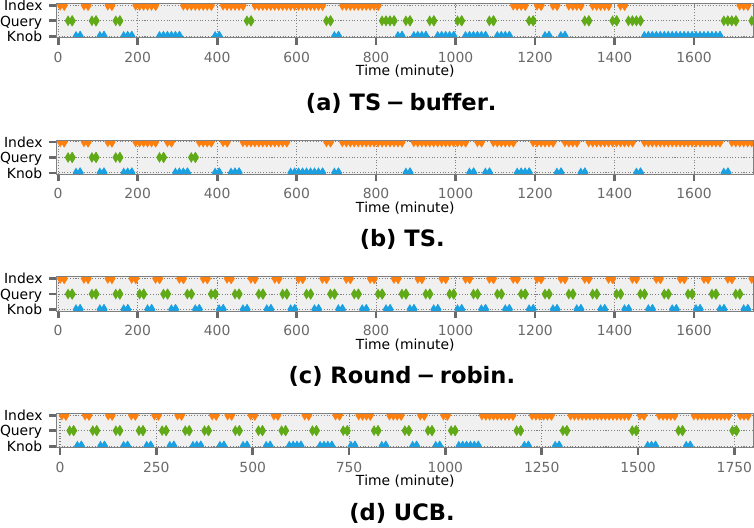}
\caption{Budget allocation under different strategies for index, knob, and query agents over time on JOB.}
\label{fig:exp-ana-budget2}
\end{figure}

We propose a Thompson Sampling based strategy with a memory buffer to allocate the tuning budgets among agents.
We compare four  allocation strategies: 
(1) TS-buffer, the Thompson Sampling  based strategy with memory buffer (the one adopted in UniTune),
(2) TS, which is the conventional  Thompson Sampling,
(3) Round-robin, which allocates the budget equally in an order of index-query-knob, as discussed in Section \ref{sec:budget},
(4) UCB, which adopts the contextual UCB algorithm~\cite{DBLP:journals/jmlr/ChuLRS11}.
It learns a mapping from context to the reward of an agent and selects the agents with the reward whose upper confidence bound is maximal.
Figure \ref{fig:exp-ana-budget1} presents the comparison result and Figure \ref{fig:exp-ana-budget2} shows their budget allocation patterns.
TS-buffer outperforms the other strategies, indicating it could allocate the tuning budget properly. 
It allocates more budget to the index agent in the first half period of tuning since configuring indexes gains more performance improvement, especially in the bootstrap phase.
And the budget  allocated to the index agent decreases in the second half period as the performance improvement of tuning indexes tends to become saturated.
TS does not dismiss any outdated information and it allocates too much budget to the index agents in the second half period.
Round-robin allocates the budget equally, wasting the budget on less promising agents (i.e., query agent).
UCB models the reward of selecting an agent in a given context. Compared with TS-buffer which directly selects the agent according to the observed rewards in a time window, learning the model requires more observations. 
Given the very limited observations (our case), UCB tends to explore the agents since the input contexts are unlikely to be covered by the previous observations.

\subsubsection{Impact of Memory Buffer Size.} \label{sec:exp-an-window}
\begin{figure}
\centering
\includegraphics{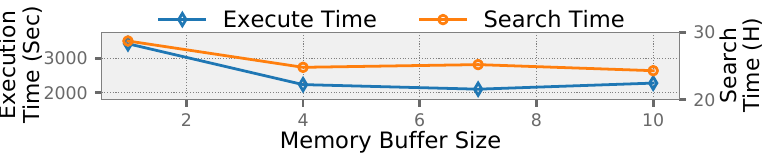}
\caption{Impact of memory buffer size. We report the best performance (i.e. minimal execution time) and the search time to reach the best performance.}
\vspace{-0.5em}
\label{fig:buffer}
\end{figure}

UniTune restricts the number of  considered  observations for the reward of selecting an agent.
While a smaller buffer size leads to a more timely response to reward changes, it may cause a loss of useful information.
We evaluate UniTune with different memory buffer sizes and present their performance in Figure \ref{fig:buffer}.
The size of one causes inferior performance since lacking the exploitation of historical observations.
In general, using buffer sizes from four to ten leads to good performance (seven is the default setting in our experiments).

\begin{table*}[]
\caption{Best performances achieved by different methods and their corresponding search time.}\label{tab:case}
\vspace{-0.5em}
\begin{tabular}{|c|c|c|c|c|c|c|c|}
\hline
                  &Default   & Grid search    & UniTune    & TS    & Round-robin & UDO   & i-q-k \\ \hline
Search Time (H)  & /   & 934.4          & \textbf{2.6} & 3.2   & 3.8         & 3.1   & 4.6   \\ \hline
Execution Time (Sec) &  581.41 & \textbf{27.11} & 27.32        & 27.32 & 27.32       & 29.87 & 29.76 \\ \hline
\end{tabular}
\end{table*}

\subsection{Case Study}\label{sec:exp-case}

We use case studies to showcase the advantage of UniTune.

\subsubsection{Comparison with Grid Search.}\label{sec:exp-case-grid}
To construct a ground-truth baseline, we define a small configuration space that can be enumerated by grid search.
We optimize one query in TPC-H, Q19.
It has 10 index candidates and 81 feasible index configurations under the 1500 MB storage budget.
We tune three important knobs\footnote{They are innodb\_buffer\_pool\_size, innodb\_log\_file\_size, innodb\_thread\_concurrency, and table\_definition\_cache, respectively.}  selected based on the Gini score and we stipulate that each knob can only have three values, resulting in 63 feasible knob configurations\footnote{The left 18 infeasible configurations violate the rule that the combined size of ib\_logfiles should be larger than 200 kB $\times$ innodb\_thread\_concurrency, causing the MySQL database to be shutdown. }.
And there are four rewrite ways for Q19, not considering loop rewriting.
To this end, we have 20412 ($81\times63\times4$) possible configurations in the configuration space, and we can evaluate their performances by a grid search to obtain the optimal configuration.
Table \ref{tab:case} presents the performance of different baselines, where we only show the best sequential tuning method due to space constraints.
Grid search finds the optimal configuration at the cost of an extremely long search time.
UniTune finds the close-to-optimal configuration with the shortest search time.
TS and round-robin can find the same configuration but with a longer search time.

\subsubsection{Comparison Between Joint Optimization and Alternating Optimization.}
\begin{figure}
\centering
\includegraphics{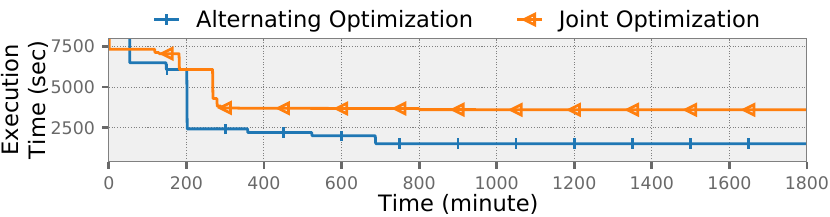}
\caption{Comparison of joint and alternating optimizations.}
\label{fig:exp-case-joint}
\end{figure}

UniTune executes the agent alternatingly.
To tune multiple components in a DBMS, another solution is to adopt a central agent over the joint configuration space, as discussed in Section \ref{sec:opt-design}.
The joint optimization has a scalability issue, due to the exponential growth of configuration space when joining the subspaces.
We validate the analysis empirically.
We adopt the two approaches to tune the knob and index components on JOB.
For alternating optimization, we adopt two BO based agents.
For joint optimization, we adopt one BO based agent optimizing over the joint configuration space.
As shown in Figure \ref{fig:exp-case-joint}, the alternating approach finds better configurations than the joint approach.
Consider the knob agent tuning 50 knobs and index agent with 59 index candidates.
For ease of analysis, we assume each knob has 10 possible values. 
Then the joint approach optimizes over a configuration space with a size of $10^{50}\times2^{59}$, $10^{17}$ times larger than the summation of the two original subspaces.
The alternating optimization addresses this scalability issue by decomposing the configuration space as the original agents assume and has a faster convergence speed.



\begin{figure}
\centering
\includegraphics{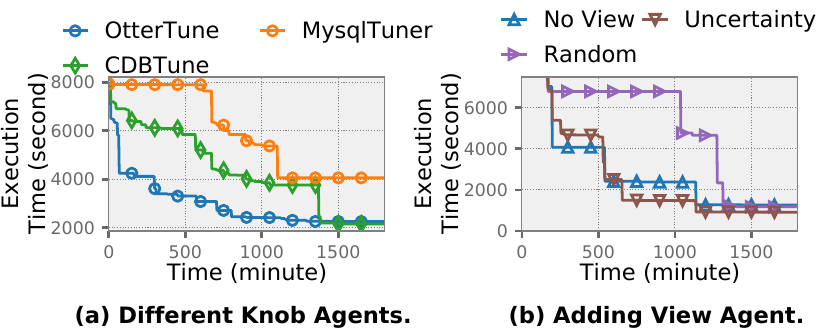}
\caption{Extendability of UniTune. In Figure 14a, we adopt three  knob agents respectively. 
In Figure 14b, we add an agent for view generation and compare the uncertainty aware sampling and random sampling discussed in Section \ref{sec:opt-rl}.}
\vspace{-0.5em}
\label{fig:exp-extend}
\end{figure}

\subsubsection{Extendability.} We integrate different tuning agents in UniTune to showcase its extendability.\label{sec:case-extend}

\textit{Adopting Different Knob Agents.} 
We still tune the three components: index, knob, and query in TPC-H. 
But, we adopt different knobs tuning agents respectively, OtterTune~\cite{DBLP:conf/sigmod/AkenPGZ17}, CDBTune~\cite{DBLP:conf/sigmod/ZhangLZLXCXWCLR19}, and MySQLTuner~\cite{MySQLTuner} in UniTune.
MySQLTuner is a rule-based tuning agent.
It examines the DBMS metrics and uses heuristic rules to suggest knob configurations.
Figure \ref{fig:exp-extend} presents the result.
UniTune converges  to similar  performances when adopting  the two ML-based tuners. 
But the convergence when adopting CDBTune is slower since its RL agent requires more tuning budget to learn a great number of neural network parameters, which is consistent with the existing studies~\cite{DBLP:journals/pvldb/AkenYBFZBP21,DBLP:journals/pvldb/ZhangCLWTLC22}.
When adopting MysqlTuner, the performance is inferior, since its limited heuristics fail to find good knob configuration.
But this adoption reveals that UniTune could be extended to non-ML based agents, such as rule-based and cost-based agents.
When a non-ML based agent  is  coordinated by UniTune, the budget allocation module treats it like the ML-based agents --
the agent executes once UniTune allocates tuning budget on it.
The main difference is that the non-ML based agents do not need the context feature to suggest configurations since they suggest configurations based on fixed rules or the cost estimation from database optimizer, which do not involve training the models.

\textit{Adding View Agent.}
We add an RL-estimator based agent Autoview~\cite{DBLP:conf/icde/Han00S21} for view generation and tune four components in JOB and set the storage budget for materialized view to 500 MB.
We compare the two strategies when training the estimator, as discussed in Section \ref{sec:opt-rl}.
As shown in Figure \ref{fig:exp-extend}b, tuning the four components archives better performance compared to tuning the three components (i.e., No View), since the materialized views save redundant computations among queries.
And the uncertainty-aware sampling achieves better performance.

\begin{figure}
\centering
\includegraphics{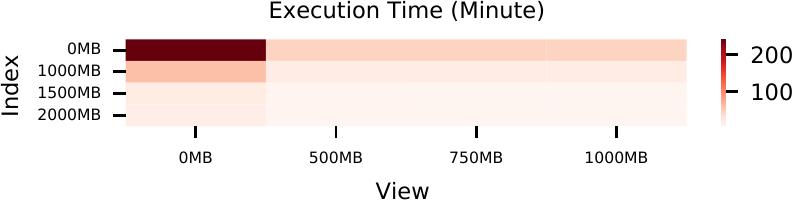}
\caption{Impact of storage budget for index and view on the best tuning performance (i.e. minimal execution time).}
\label{fig:exp-space}
\vspace{-1em}
\end{figure}

\subsubsection{Varying Storage Budgets}\label{exp:case-space}
At the start of a tuning session, the storage budget is decided by users based on their affordable resource.
In this experiment, we vary the storage budgets for index and view on JOB workload and present the best performance tuned by UniTune in Figure \ref{fig:exp-space}.
We observe that the tuning performance gradually increases with the given storage budget and then saturates. For the view component, its tuning benefit saturates after being given 500MB storage budget, which is consistent with AutoView's report ~\cite{DBLP:conf/icde/Han00S21}.
For the index component, its tuning benefit increases significantly from zero storage budget to 1500 MB and then the increase becomes gentle.
In our experiments, we use 1500MB for index storage budget and 500MB for view storage budget on the JOB workload.

\section{Conclusion}
In this paper, we reviewed the emerging studies on  ML-based tuning agents in the database community and raised a question --  ``how to make them work together to configure multiple components  in a DBMS?''
To answer this, we proposed a unified and efficient coordinating framework UniTune for the ML-based tuning agents.
We design a coordination protocol to enable collaboration among tuning agents and a strategy to allocate the tuning budget among the agents.
We also define the interfaces adapted to a broad
class of ML-based tuning agents, which are simple for integration with existing implementation and for future extension.
We demonstrated that UniTune could support different ML-based agents and significantly outperforms the baselines.

\begin{acks}
This work is supported by National Natural Science Foundation of China (NSFC)(No. 61832001, U22B2037), Alibaba Group through Alibaba Innovative Research Program and National Key Research. 
Bin Cui and Yang Li are the corresponding author.
\end{acks}

\balance
\bibliographystyle{ACM-Reference-Format}
\bibliography{sample-base}


\begin{thebibliography}{74}


\ifx \showCODEN    \undefined \def \showCODEN     #1{\unskip}     \fi
\ifx \showDOI      \undefined \def \showDOI       #1{#1}\fi
\ifx \showISBNx    \undefined \def \showISBNx     #1{\unskip}     \fi
\ifx \showISBNxiii \undefined \def \showISBNxiii  #1{\unskip}     \fi
\ifx \showISSN     \undefined \def \showISSN      #1{\unskip}     \fi
\ifx \showLCCN     \undefined \def \showLCCN      #1{\unskip}     \fi
\ifx \shownote     \undefined \def \shownote      #1{#1}          \fi
\ifx \showarticletitle \undefined \def \showarticletitle #1{#1}   \fi
\ifx \showURL      \undefined \def \showURL       {\relax}        \fi
\providecommand\bibfield[2]{#2}
\providecommand\bibinfo[2]{#2}
\providecommand\natexlab[1]{#1}
\providecommand\showeprint[2][]{arXiv:#2}

\bibitem[tpc(2015a)]%
        {tpch}
 \bibinfo{year}{2015}\natexlab{a}.
\newblock \bibinfo{title}{TPC-H benchmark}.
\newblock \bibinfo{howpublished}{\url{http://www.tpc.org/tpch/}}.
\newblock


\bibitem[tpc(2015b)]%
        {tpcds}
 \bibinfo{year}{2015}\natexlab{b}.
\newblock \bibinfo{title}{TPC-H benchmark}.
\newblock \bibinfo{howpublished}{\url{https://www.tpc.org/tpcds/}}.
\newblock


\bibitem[udo(2015)]%
        {udo}
 \bibinfo{year}{2015}\natexlab{}.
\newblock \bibinfo{title}{UDO Implementation}.
\newblock \bibinfo{howpublished}{\url{https://github.com/jxiw/UDO}}.
\newblock


\bibitem[MyS(2019)]%
        {MySQLTuner}
 \bibinfo{year}{2019}\natexlab{}.
\newblock \bibinfo{title}{MySQL Tuning Primer Script}.
\newblock
  \bibinfo{howpublished}{\url{https://github.com/major/MySQLTuner-perl}}.
\newblock


\bibitem[Agrawal and Goyal(2012)]%
        {DBLP:journals/jmlr/AgrawalG12}
\bibfield{author}{\bibinfo{person}{Shipra Agrawal} {and} \bibinfo{person}{Navin
  Goyal}.} \bibinfo{year}{2012}\natexlab{}.
\newblock \showarticletitle{Analysis of Thompson Sampling for the Multi-armed
  Bandit Problem}. In \bibinfo{booktitle}{\emph{{COLT}}}
  \emph{(\bibinfo{series}{{JMLR} Proceedings}, Vol.~\bibinfo{volume}{23})}.
  \bibinfo{publisher}{JMLR.org}, \bibinfo{pages}{39.1--39.26}.
\newblock


\bibitem[Aken et~al\mbox{.}(2017)]%
        {DBLP:conf/sigmod/AkenPGZ17}
\bibfield{author}{\bibinfo{person}{Dana~Van Aken}, \bibinfo{person}{Andrew
  Pavlo}, \bibinfo{person}{Geoffrey~J. Gordon}, {and} \bibinfo{person}{Bohan
  Zhang}.} \bibinfo{year}{2017}\natexlab{}.
\newblock \showarticletitle{Automatic Database Management System Tuning Through
  Large-scale Machine Learning}. In \bibinfo{booktitle}{\emph{{SIGMOD}
  Conference}}. \bibinfo{publisher}{{ACM}}, \bibinfo{pages}{1009--1024}.
\newblock


\bibitem[Aken et~al\mbox{.}(2021)]%
        {DBLP:journals/pvldb/AkenYBFZBP21}
\bibfield{author}{\bibinfo{person}{Dana~Van Aken}, \bibinfo{person}{Dongsheng
  Yang}, \bibinfo{person}{Sebastien Brillard}, \bibinfo{person}{Ari Fiorino},
  \bibinfo{person}{Bohan Zhang}, \bibinfo{person}{Christian Billian}, {and}
  \bibinfo{person}{Andrew Pavlo}.} \bibinfo{year}{2021}\natexlab{}.
\newblock \showarticletitle{An Inquiry into Machine Learning-based Automatic
  Configuration Tuning Services on Real-World Database Management Systems}.
\newblock \bibinfo{journal}{\emph{Proc. {VLDB} Endow.}} \bibinfo{volume}{14},
  \bibinfo{number}{7} (\bibinfo{year}{2021}), \bibinfo{pages}{1241--1253}.
\newblock


\bibitem[Begoli et~al\mbox{.}(2018)]%
        {DBLP:conf/sigmod/BegoliCHML18}
\bibfield{author}{\bibinfo{person}{Edmon Begoli}, \bibinfo{person}{Jes{\'{u}}s
  Camacho{-}Rodr{\'{\i}}guez}, \bibinfo{person}{Julian Hyde},
  \bibinfo{person}{Michael~J. Mior}, {and} \bibinfo{person}{Daniel Lemire}.}
  \bibinfo{year}{2018}\natexlab{}.
\newblock \showarticletitle{Apache Calcite: {A} Foundational Framework for
  Optimized Query Processing Over Heterogeneous Data Sources}. In
  \bibinfo{booktitle}{\emph{{SIGMOD} Conference}}. \bibinfo{publisher}{{ACM}},
  \bibinfo{pages}{221--230}.
\newblock


\bibitem[Bubeck and Cesa{-}Bianchi(2012)]%
        {DBLP:journals/ftml/BubeckC12}
\bibfield{author}{\bibinfo{person}{S{\'{e}}bastien Bubeck} {and}
  \bibinfo{person}{Nicol{\`{o}} Cesa{-}Bianchi}.}
  \bibinfo{year}{2012}\natexlab{}.
\newblock \showarticletitle{Regret Analysis of Stochastic and Nonstochastic
  Multi-armed Bandit Problems}.
\newblock \bibinfo{journal}{\emph{Found. Trends Mach. Learn.}}
  \bibinfo{volume}{5}, \bibinfo{number}{1} (\bibinfo{year}{2012}),
  \bibinfo{pages}{1--122}.
\newblock


\bibitem[Cai et~al\mbox{.}(2022)]%
        {DBLP:conf/sigmod/CaiLZZZLLCYX22}
\bibfield{author}{\bibinfo{person}{Baoqing Cai}, \bibinfo{person}{Yu Liu},
  \bibinfo{person}{Ce Zhang}, \bibinfo{person}{Guangyu Zhang},
  \bibinfo{person}{Ke Zhou}, \bibinfo{person}{Li Liu}, \bibinfo{person}{Chunhua
  Li}, \bibinfo{person}{Bin Cheng}, \bibinfo{person}{Jie Yang}, {and}
  \bibinfo{person}{Jiashu Xing}.} \bibinfo{year}{2022}\natexlab{}.
\newblock \showarticletitle{{HUNTER:} An Online Cloud Database Hybrid Tuning
  System for Personalized Requirements}. In \bibinfo{booktitle}{\emph{{SIGMOD}
  Conference}}. \bibinfo{publisher}{{ACM}}, \bibinfo{pages}{646--659}.
\newblock


\bibitem[Cereda et~al\mbox{.}(2021)]%
        {DBLP:journals/pvldb/CeredaVCD21}
\bibfield{author}{\bibinfo{person}{Stefano Cereda}, \bibinfo{person}{Stefano
  Valladares}, \bibinfo{person}{Paolo Cremonesi}, {and}
  \bibinfo{person}{Stefano Doni}.} \bibinfo{year}{2021}\natexlab{}.
\newblock \showarticletitle{CGPTuner: a Contextual Gaussian Process Bandit
  Approach for the Automatic Tuning of {IT} Configurations Under Varying
  Workload Conditions}.
\newblock \bibinfo{journal}{\emph{Proc. {VLDB} Endow.}} \bibinfo{volume}{14},
  \bibinfo{number}{8} (\bibinfo{year}{2021}), \bibinfo{pages}{1401--1413}.
\newblock


\bibitem[Chaudhuri and Narasayya(1997)]%
        {DBLP:conf/vldb/ChaudhuriN97}
\bibfield{author}{\bibinfo{person}{Surajit Chaudhuri} {and}
  \bibinfo{person}{Vivek~R. Narasayya}.} \bibinfo{year}{1997}\natexlab{}.
\newblock \showarticletitle{An Efficient Cost-Driven Index Selection Tool for
  Microsoft {SQL} Server}. In \bibinfo{booktitle}{\emph{VLDB'97, Proceedings of
  23rd International Conference on Very Large Data Bases, August 25-29, 1997,
  Athens, Greece}}, \bibfield{editor}{\bibinfo{person}{Matthias Jarke},
  \bibinfo{person}{Michael~J. Carey}, \bibinfo{person}{Klaus~R. Dittrich},
  \bibinfo{person}{Frederick~H. Lochovsky}, \bibinfo{person}{Pericles
  Loucopoulos}, {and} \bibinfo{person}{Manfred~A. Jeusfeld}} (Eds.).
  \bibinfo{publisher}{Morgan Kaufmann}, \bibinfo{pages}{146--155}.
\newblock
\urldef\tempurl%
\url{http://www.vldb.org/conf/1997/P146.PDF}
\showURL{%
\tempurl}


\bibitem[Chen et~al\mbox{.}(2022)]%
        {DBLP:conf/kdd/0008Y000CZ022}
\bibfield{author}{\bibinfo{person}{Jin Chen}, \bibinfo{person}{Guanyu Ye},
  \bibinfo{person}{Yan Zhao}, \bibinfo{person}{Shuncheng Liu},
  \bibinfo{person}{Liwei Deng}, \bibinfo{person}{Xu Chen}, \bibinfo{person}{Rui
  Zhou}, {and} \bibinfo{person}{Kai Zheng}.} \bibinfo{year}{2022}\natexlab{}.
\newblock \showarticletitle{Efficient Join Order Selection Learning with
  Graph-based Representation}. In \bibinfo{booktitle}{\emph{{KDD}}}.
  \bibinfo{publisher}{{ACM}}, \bibinfo{pages}{97--107}.
\newblock


\bibitem[Chu et~al\mbox{.}(2011)]%
        {DBLP:journals/jmlr/ChuLRS11}
\bibfield{author}{\bibinfo{person}{Wei Chu}, \bibinfo{person}{Lihong Li},
  \bibinfo{person}{Lev Reyzin}, {and} \bibinfo{person}{Robert~E. Schapire}.}
  \bibinfo{year}{2011}\natexlab{}.
\newblock \showarticletitle{Contextual Bandits with Linear Payoff Functions}.
  In \bibinfo{booktitle}{\emph{{AISTATS}}} \emph{(\bibinfo{series}{{JMLR}
  Proceedings}, Vol.~\bibinfo{volume}{15})}. \bibinfo{publisher}{JMLR.org},
  \bibinfo{pages}{208--214}.
\newblock


\bibitem[Ding et~al\mbox{.}(2019)]%
        {DBLP:conf/sigmod/DingDM0CN19}
\bibfield{author}{\bibinfo{person}{Bailu Ding}, \bibinfo{person}{Sudipto Das},
  \bibinfo{person}{Ryan Marcus}, \bibinfo{person}{Wentao Wu},
  \bibinfo{person}{Surajit Chaudhuri}, {and} \bibinfo{person}{Vivek~R.
  Narasayya}.} \bibinfo{year}{2019}\natexlab{}.
\newblock \showarticletitle{{AI} Meets {AI:} Leveraging Query Executions to
  Improve Index Recommendations}. In \bibinfo{booktitle}{\emph{{SIGMOD}
  Conference}}. \bibinfo{publisher}{{ACM}}, \bibinfo{pages}{1241--1258}.
\newblock


\bibitem[Duan et~al\mbox{.}(2009)]%
        {DBLP:journals/pvldb/DuanTB09}
\bibfield{author}{\bibinfo{person}{Songyun Duan}, \bibinfo{person}{Vamsidhar
  Thummala}, {and} \bibinfo{person}{Shivnath Babu}.}
  \bibinfo{year}{2009}\natexlab{}.
\newblock \showarticletitle{Tuning Database Configuration Parameters with
  iTuned}.
\newblock \bibinfo{journal}{\emph{Proc. {VLDB} Endow.}} \bibinfo{volume}{2},
  \bibinfo{number}{1} (\bibinfo{year}{2009}), \bibinfo{pages}{1246--1257}.
\newblock


\bibitem[Durand et~al\mbox{.}(2018)]%
        {DBLP:conf/sigmod/DurandPPMBSSRB18}
\bibfield{author}{\bibinfo{person}{Gabriel~Campero Durand},
  \bibinfo{person}{Marcus Pinnecke}, \bibinfo{person}{Rufat Piriyev},
  \bibinfo{person}{Mahmoud Mohsen}, \bibinfo{person}{David Broneske},
  \bibinfo{person}{Gunter Saake}, \bibinfo{person}{Maya~S. Sekeran},
  \bibinfo{person}{Fabi{\'{a}}n Rodriguez}, {and} \bibinfo{person}{Laxmi
  Balami}.} \bibinfo{year}{2018}\natexlab{}.
\newblock \showarticletitle{GridFormation: Towards Self-Driven Online Data
  Partitioning using Reinforcement Learning}. In
  \bibinfo{booktitle}{\emph{aiDM@SIGMOD}}. \bibinfo{publisher}{{ACM}},
  \bibinfo{pages}{1:1--1:7}.
\newblock


\bibitem[Durand et~al\mbox{.}(2019)]%
        {DBLP:conf/adbis/DurandPPBGS19}
\bibfield{author}{\bibinfo{person}{Gabriel~Campero Durand},
  \bibinfo{person}{Rufat Piriyev}, \bibinfo{person}{Marcus Pinnecke},
  \bibinfo{person}{David Broneske}, \bibinfo{person}{Balasubramanian
  Gurumurthy}, {and} \bibinfo{person}{Gunter Saake}.}
  \bibinfo{year}{2019}\natexlab{}.
\newblock \showarticletitle{Automated Vertical Partitioning with Deep
  Reinforcement Learning}. In \bibinfo{booktitle}{\emph{{ADBIS} (Short Papers
  and Workshops)}} \emph{(\bibinfo{series}{Communications in Computer and
  Information Science}, Vol.~\bibinfo{volume}{1064})}.
  \bibinfo{publisher}{Springer}, \bibinfo{pages}{126--134}.
\newblock


\bibitem[Eldeeb et~al\mbox{.}(2022)]%
        {DBLP:conf/sigmod/EldeebCCY22}
\bibfield{author}{\bibinfo{person}{Tamer Eldeeb}, \bibinfo{person}{Zhengneng
  Chen}, \bibinfo{person}{Asaf Cidon}, {and} \bibinfo{person}{Junfeng Yang}.}
  \bibinfo{year}{2022}\natexlab{}.
\newblock \showarticletitle{Neuroshard: towards automatic multi-objective
  sharding with deep reinforcement learning}. In
  \bibinfo{booktitle}{\emph{aiDM@SIGMOD}}. \bibinfo{publisher}{{ACM}},
  \bibinfo{pages}{1:1--1:12}.
\newblock


\bibitem[Fekry et~al\mbox{.}(2020)]%
        {DBLP:conf/kdd/FekryCPRH20}
\bibfield{author}{\bibinfo{person}{Ayat Fekry}, \bibinfo{person}{Lucian
  Carata}, \bibinfo{person}{Thomas F.~J.{-}M. Pasquier},
  \bibinfo{person}{Andrew Rice}, {and} \bibinfo{person}{Andy Hopper}.}
  \bibinfo{year}{2020}\natexlab{}.
\newblock \showarticletitle{To Tune or Not to Tune?: In Search of Optimal
  Configurations for Data Analytics}. In \bibinfo{booktitle}{\emph{{KDD}}}.
  \bibinfo{publisher}{{ACM}}, \bibinfo{pages}{2494--2504}.
\newblock


\bibitem[Finance and Gardarin(1991)]%
        {DBLP:conf/icde/FinanceG91}
\bibfield{author}{\bibinfo{person}{B{\'{e}}atrice Finance} {and}
  \bibinfo{person}{Georges Gardarin}.} \bibinfo{year}{1991}\natexlab{}.
\newblock \showarticletitle{A Rule-Based Query Rewriter in an Extensible
  {DBMS}}. In \bibinfo{booktitle}{\emph{{ICDE}}}. \bibinfo{publisher}{{IEEE}
  Computer Society}, \bibinfo{pages}{248--256}.
\newblock


\bibitem[Gao et~al\mbox{.}(2022)]%
        {Gao2022AutomaticIS}
\bibfield{author}{\bibinfo{person}{Jianling Gao}, \bibinfo{person}{Nan Zhao},
  \bibinfo{person}{Ning Wang}, \bibinfo{person}{Shuang Hao}, {and}
  \bibinfo{person}{Haoyan Wu}.} \bibinfo{year}{2022}\natexlab{}.
\newblock \showarticletitle{Automatic Index Selection with Learned Cost
  Estimator}.
\newblock \bibinfo{journal}{\emph{Information Sciences}}
  (\bibinfo{year}{2022}).
\newblock


\bibitem[Ge et~al\mbox{.}(2021)]%
        {DBLP:journals/jcst/GeCC21}
\bibfield{author}{\bibinfo{person}{Jia{-}Ke Ge}, \bibinfo{person}{Yanfeng
  Chai}, {and} \bibinfo{person}{Yunpeng Chai}.}
  \bibinfo{year}{2021}\natexlab{}.
\newblock \showarticletitle{WATuning: {A} Workload-Aware Tuning System with
  Attention-Based Deep Reinforcement Learning}.
\newblock \bibinfo{journal}{\emph{J. Comput. Sci. Technol.}}
  \bibinfo{volume}{36}, \bibinfo{number}{4} (\bibinfo{year}{2021}),
  \bibinfo{pages}{741--761}.
\newblock


\bibitem[Gur et~al\mbox{.}(2021)]%
        {DBLP:conf/edbt/GurYSR21}
\bibfield{author}{\bibinfo{person}{Yaniv Gur}, \bibinfo{person}{Dongsheng
  Yang}, \bibinfo{person}{Frederik Stalschus}, {and} \bibinfo{person}{Berthold
  Reinwald}.} \bibinfo{year}{2021}\natexlab{}.
\newblock \showarticletitle{Adaptive Multi-Model Reinforcement Learning for
  Online Database Tuning}. In \bibinfo{booktitle}{\emph{{EDBT}}}.
  \bibinfo{publisher}{OpenProceedings.org}, \bibinfo{pages}{439--444}.
\newblock


\bibitem[Gur et~al\mbox{.}(2014)]%
        {DBLP:conf/nips/GurZB14}
\bibfield{author}{\bibinfo{person}{Yonatan Gur}, \bibinfo{person}{Assaf Zeevi},
  {and} \bibinfo{person}{Omar Besbes}.} \bibinfo{year}{2014}\natexlab{}.
\newblock \showarticletitle{Stochastic Multi-Armed-Bandit Problem with
  Non-stationary Rewards}. In \bibinfo{booktitle}{\emph{{NIPS}}}.
  \bibinfo{pages}{199--207}.
\newblock


\bibitem[Han et~al\mbox{.}(2022)]%
        {DBLP:journals/jcst/HanLL22}
\bibfield{author}{\bibinfo{person}{Shuai Han}, \bibinfo{person}{Mingxia Liu},
  {and} \bibinfo{person}{Jian{-}Zhong Li}.} \bibinfo{year}{2022}\natexlab{}.
\newblock \showarticletitle{Efficient Partitioning Method for Optimizing the
  Compression on Array Data}.
\newblock \bibinfo{journal}{\emph{J. Comput. Sci. Technol.}}
  \bibinfo{volume}{37}, \bibinfo{number}{5} (\bibinfo{year}{2022}),
  \bibinfo{pages}{1049--1067}.
\newblock


\bibitem[Han et~al\mbox{.}(2021)]%
        {DBLP:conf/icde/Han00S21}
\bibfield{author}{\bibinfo{person}{Yue Han}, \bibinfo{person}{Guoliang Li},
  \bibinfo{person}{Haitao Yuan}, {and} \bibinfo{person}{Ji Sun}.}
  \bibinfo{year}{2021}\natexlab{}.
\newblock \showarticletitle{An Autonomous Materialized View Management System
  with Deep Reinforcement Learning}. In \bibinfo{booktitle}{\emph{{ICDE}}}.
  \bibinfo{publisher}{{IEEE}}, \bibinfo{pages}{2159--2164}.
\newblock


\bibitem[Heitz and Stockinger(2019)]%
        {DBLP:journals/corr/abs-1911-11689}
\bibfield{author}{\bibinfo{person}{Jonas Heitz} {and} \bibinfo{person}{Kurt
  Stockinger}.} \bibinfo{year}{2019}\natexlab{}.
\newblock \showarticletitle{Join Query Optimization with Deep Reinforcement
  Learning Algorithms}.
\newblock \bibinfo{journal}{\emph{CoRR}}  \bibinfo{volume}{abs/1911.11689}
  (\bibinfo{year}{2019}).
\newblock


\bibitem[Hilprecht et~al\mbox{.}(2020)]%
        {DBLP:conf/sigmod/HilprechtBR20}
\bibfield{author}{\bibinfo{person}{Benjamin Hilprecht},
  \bibinfo{person}{Carsten Binnig}, {and} \bibinfo{person}{Uwe R{\"{o}}hm}.}
  \bibinfo{year}{2020}\natexlab{}.
\newblock \showarticletitle{Learning a Partitioning Advisor for Cloud
  Databases}. In \bibinfo{booktitle}{\emph{{SIGMOD} Conference}}.
  \bibinfo{publisher}{{ACM}}, \bibinfo{pages}{143--157}.
\newblock


\bibitem[Huang et~al\mbox{.}(2023)]%
        {DBLP:journals/chinaf/HuangQZTLC23}
\bibfield{author}{\bibinfo{person}{Shiyue Huang}, \bibinfo{person}{Yanzhao
  Qin}, \bibinfo{person}{Xinyi Zhang}, \bibinfo{person}{Yaofeng Tu},
  \bibinfo{person}{Zhongliang Li}, {and} \bibinfo{person}{Bin Cui}.}
  \bibinfo{year}{2023}\natexlab{}.
\newblock \showarticletitle{Survey on performance optimization for database
  systems}.
\newblock \bibinfo{journal}{\emph{Sci. China Inf. Sci.}} \bibinfo{volume}{66},
  \bibinfo{number}{2} (\bibinfo{year}{2023}).
\newblock


\bibitem[Ishihara and Shiba(2020)]%
        {DBLP:conf/lifetech/IshiharaS20}
\bibfield{author}{\bibinfo{person}{Yoshiteru Ishihara} {and}
  \bibinfo{person}{Masahito Shiba}.} \bibinfo{year}{2020}\natexlab{}.
\newblock \showarticletitle{Dynamic Configuration Tuning of Working Database
  Management Systems}. In \bibinfo{booktitle}{\emph{LifeTech}}.
  \bibinfo{publisher}{{IEEE}}, \bibinfo{pages}{393--397}.
\newblock


\bibitem[Kanellis et~al\mbox{.}(2020)]%
        {DBLP:conf/hotstorage/KanellisAV20}
\bibfield{author}{\bibinfo{person}{Konstantinos Kanellis},
  \bibinfo{person}{Ramnatthan Alagappan}, {and} \bibinfo{person}{Shivaram
  Venkataraman}.} \bibinfo{year}{2020}\natexlab{}.
\newblock \showarticletitle{Too Many Knobs to Tune? Towards Faster Database
  Tuning by Pre-selecting Important Knobs}. In
  \bibinfo{booktitle}{\emph{HotStorage}}. \bibinfo{publisher}{{USENIX}
  Association}.
\newblock


\bibitem[Kanellis et~al\mbox{.}(2022)]%
        {DBLP:journals/pvldb/KanellisDKMCV22}
\bibfield{author}{\bibinfo{person}{Konstantinos Kanellis},
  \bibinfo{person}{Cong Ding}, \bibinfo{person}{Brian Kroth},
  \bibinfo{person}{Andreas M{\"{u}}ller}, \bibinfo{person}{Carlo Curino}, {and}
  \bibinfo{person}{Shivaram Venkataraman}.} \bibinfo{year}{2022}\natexlab{}.
\newblock \showarticletitle{LlamaTune: Sample-Efficient {DBMS} Configuration
  Tuning}.
\newblock \bibinfo{journal}{\emph{Proc. {VLDB} Endow.}} \bibinfo{volume}{15},
  \bibinfo{number}{11} (\bibinfo{year}{2022}), \bibinfo{pages}{2953--2965}.
\newblock


\bibitem[Kossmann et~al\mbox{.}(2020)]%
        {DBLP:journals/pvldb/KossmannHJS20}
\bibfield{author}{\bibinfo{person}{Jan Kossmann}, \bibinfo{person}{Stefan
  Halfpap}, \bibinfo{person}{Marcel Jankrift}, {and} \bibinfo{person}{Rainer
  Schlosser}.} \bibinfo{year}{2020}\natexlab{}.
\newblock \showarticletitle{Magic mirror in my hand, which is the best in the
  land? An Experimental Evaluation of Index Selection Algorithms}.
\newblock \bibinfo{journal}{\emph{Proc. {VLDB} Endow.}} \bibinfo{volume}{13},
  \bibinfo{number}{11} (\bibinfo{year}{2020}), \bibinfo{pages}{2382--2395}.
\newblock


\bibitem[Kossmann et~al\mbox{.}(2022)]%
        {DBLP:conf/edbt/KossmannKS22}
\bibfield{author}{\bibinfo{person}{Jan Kossmann}, \bibinfo{person}{Alexander
  Kastius}, {and} \bibinfo{person}{Rainer Schlosser}.}
  \bibinfo{year}{2022}\natexlab{}.
\newblock \showarticletitle{{SWIRL:} Selection of Workload-aware Indexes using
  Reinforcement Learning}. In \bibinfo{booktitle}{\emph{{EDBT}}}.
  \bibinfo{publisher}{OpenProceedings.org}, \bibinfo{pages}{2:155--2:168}.
\newblock


\bibitem[Krishnan et~al\mbox{.}(2018)]%
        {DBLP:journals/corr/abs-1808-03196}
\bibfield{author}{\bibinfo{person}{Sanjay Krishnan}, \bibinfo{person}{Zongheng
  Yang}, \bibinfo{person}{Ken Goldberg}, \bibinfo{person}{Joseph~M.
  Hellerstein}, {and} \bibinfo{person}{Ion Stoica}.}
  \bibinfo{year}{2018}\natexlab{}.
\newblock \showarticletitle{Learning to Optimize Join Queries With Deep
  Reinforcement Learning}.
\newblock \bibinfo{journal}{\emph{CoRR}}  \bibinfo{volume}{abs/1808.03196}
  (\bibinfo{year}{2018}).
\newblock


\bibitem[Kunjir and Babu(2020)]%
        {DBLP:conf/sigmod/KunjirB20}
\bibfield{author}{\bibinfo{person}{Mayuresh Kunjir} {and}
  \bibinfo{person}{Shivnath Babu}.} \bibinfo{year}{2020}\natexlab{}.
\newblock \showarticletitle{Black or White? How to Develop an AutoTuner for
  Memory-based Analytics}. In \bibinfo{booktitle}{\emph{{SIGMOD} Conference}}.
  \bibinfo{publisher}{{ACM}}, \bibinfo{pages}{1667--1683}.
\newblock


\bibitem[Lan et~al\mbox{.}(2020)]%
        {DBLP:conf/cikm/LanBP20}
\bibfield{author}{\bibinfo{person}{Hai Lan}, \bibinfo{person}{Zhifeng Bao},
  {and} \bibinfo{person}{Yuwei Peng}.} \bibinfo{year}{2020}\natexlab{}.
\newblock \showarticletitle{An Index Advisor Using Deep Reinforcement
  Learning}. In \bibinfo{booktitle}{\emph{{CIKM}}}. \bibinfo{publisher}{{ACM}},
  \bibinfo{pages}{2105--2108}.
\newblock


\bibitem[Lan et~al\mbox{.}(2021)]%
        {DBLP:journals/dase/LanBP21}
\bibfield{author}{\bibinfo{person}{Hai Lan}, \bibinfo{person}{Zhifeng Bao},
  {and} \bibinfo{person}{Yuwei Peng}.} \bibinfo{year}{2021}\natexlab{}.
\newblock \showarticletitle{A Survey on Advancing the {DBMS} Query Optimizer:
  Cardinality Estimation, Cost Model, and Plan Enumeration}.
\newblock \bibinfo{journal}{\emph{Data Sci. Eng.}} \bibinfo{volume}{6},
  \bibinfo{number}{1} (\bibinfo{year}{2021}), \bibinfo{pages}{86--101}.
\newblock


\bibitem[Leis et~al\mbox{.}(2015)]%
        {DBLP:journals/pvldb/LeisGMBK015}
\bibfield{author}{\bibinfo{person}{Viktor Leis}, \bibinfo{person}{Andrey
  Gubichev}, \bibinfo{person}{Atanas Mirchev}, \bibinfo{person}{Peter~A.
  Boncz}, \bibinfo{person}{Alfons Kemper}, {and} \bibinfo{person}{Thomas
  Neumann}.} \bibinfo{year}{2015}\natexlab{}.
\newblock \showarticletitle{How Good Are Query Optimizers, Really?}
\newblock \bibinfo{journal}{\emph{Proc. {VLDB} Endow.}} \bibinfo{volume}{9},
  \bibinfo{number}{3} (\bibinfo{year}{2015}), \bibinfo{pages}{204--215}.
\newblock


\bibitem[Li et~al\mbox{.}(2021b)]%
        {DBLP:journals/pvldb/0001Z021}
\bibfield{author}{\bibinfo{person}{Guoliang Li}, \bibinfo{person}{Xuanhe Zhou},
  {and} \bibinfo{person}{Lei Cao}.} \bibinfo{year}{2021}\natexlab{b}.
\newblock \showarticletitle{Machine Learning for Databases}.
\newblock \bibinfo{journal}{\emph{Proc. {VLDB} Endow.}} \bibinfo{volume}{14},
  \bibinfo{number}{12} (\bibinfo{year}{2021}), \bibinfo{pages}{3190--3193}.
\newblock


\bibitem[Li et~al\mbox{.}(2019)]%
        {DBLP:journals/pvldb/LiZLG19}
\bibfield{author}{\bibinfo{person}{Guoliang Li}, \bibinfo{person}{Xuanhe Zhou},
  \bibinfo{person}{Shifu Li}, {and} \bibinfo{person}{Bo Gao}.}
  \bibinfo{year}{2019}\natexlab{}.
\newblock \showarticletitle{QTune: {A} Query-Aware Database Tuning System with
  Deep Reinforcement Learning}.
\newblock \bibinfo{journal}{\emph{Proc. {VLDB} Endow.}} \bibinfo{volume}{12},
  \bibinfo{number}{12} (\bibinfo{year}{2019}), \bibinfo{pages}{2118--2130}.
\newblock


\bibitem[Li et~al\mbox{.}(2020)]%
        {DBLP:conf/aaai/LiJGSZ020}
\bibfield{author}{\bibinfo{person}{Yang Li}, \bibinfo{person}{Jiawei Jiang},
  \bibinfo{person}{Jinyang Gao}, \bibinfo{person}{Yingxia Shao},
  \bibinfo{person}{Ce Zhang}, {and} \bibinfo{person}{Bin Cui}.}
  \bibinfo{year}{2020}\natexlab{}.
\newblock \showarticletitle{Efficient Automatic {CASH} via Rising Bandits}. In
  \bibinfo{booktitle}{\emph{{AAAI}}}. \bibinfo{publisher}{{AAAI} Press},
  \bibinfo{pages}{4763--4771}.
\newblock


\bibitem[Li et~al\mbox{.}(2022)]%
        {DBLP:conf/kdd/LiSJBZZC22}
\bibfield{author}{\bibinfo{person}{Yang Li}, \bibinfo{person}{Yu Shen},
  \bibinfo{person}{Huaijun Jiang}, \bibinfo{person}{Tianyi Bai},
  \bibinfo{person}{Wentao Zhang}, \bibinfo{person}{Ce Zhang}, {and}
  \bibinfo{person}{Bin Cui}.} \bibinfo{year}{2022}\natexlab{}.
\newblock \showarticletitle{Transfer Learning based Search Space Design for
  Hyperparameter Tuning}. In \bibinfo{booktitle}{\emph{{KDD}}}.
  \bibinfo{publisher}{{ACM}}, \bibinfo{pages}{967--977}.
\newblock


\bibitem[Li et~al\mbox{.}(2021a)]%
        {DBLP:journals/pvldb/LiSZJLDZY00021}
\bibfield{author}{\bibinfo{person}{Yang Li}, \bibinfo{person}{Yu Shen},
  \bibinfo{person}{Wentao Zhang}, \bibinfo{person}{Jiawei Jiang},
  \bibinfo{person}{Yaliang Li}, \bibinfo{person}{Bolin Ding},
  \bibinfo{person}{Jingren Zhou}, \bibinfo{person}{Zhi Yang},
  \bibinfo{person}{Wentao Wu}, \bibinfo{person}{Ce Zhang}, {and}
  \bibinfo{person}{Bin Cui}.} \bibinfo{year}{2021}\natexlab{a}.
\newblock \showarticletitle{VolcanoML: Speeding up End-to-End AutoML via
  Scalable Search Space Decomposition}.
\newblock \bibinfo{journal}{\emph{Proc. {VLDB} Endow.}} \bibinfo{volume}{14},
  \bibinfo{number}{11} (\bibinfo{year}{2021}), \bibinfo{pages}{2167--2176}.
\newblock


\bibitem[Liang et~al\mbox{.}(2019)]%
        {DBLP:journals/corr/abs-1903-01363}
\bibfield{author}{\bibinfo{person}{Xi Liang}, \bibinfo{person}{Aaron~J.
  Elmore}, {and} \bibinfo{person}{Sanjay Krishnan}.}
  \bibinfo{year}{2019}\natexlab{}.
\newblock \showarticletitle{Opportunistic View Materialization with Deep
  Reinforcement Learning}.
\newblock \bibinfo{journal}{\emph{CoRR}}  \bibinfo{volume}{abs/1903.01363}
  (\bibinfo{year}{2019}).
\newblock


\bibitem[Licks et~al\mbox{.}(2020)]%
        {DBLP:journals/apin/LicksCMPRM20}
\bibfield{author}{\bibinfo{person}{Gabriel~Paludo Licks},
  \bibinfo{person}{J{\'{u}}lia Mara~Colleoni Couto}, \bibinfo{person}{Priscilla
  de F{\'{a}}tima~Miehe}, \bibinfo{person}{Renata~De Paris},
  \bibinfo{person}{Duncan Dubugras~A. Ruiz}, {and} \bibinfo{person}{Felipe
  Meneguzzi}.} \bibinfo{year}{2020}\natexlab{}.
\newblock \showarticletitle{SmartIX: {A} database indexing agent based on
  reinforcement learning}.
\newblock \bibinfo{journal}{\emph{Appl. Intell.}} \bibinfo{volume}{50},
  \bibinfo{number}{8} (\bibinfo{year}{2020}), \bibinfo{pages}{2575--2588}.
\newblock


\bibitem[Liu et~al\mbox{.}(2020)]%
        {DBLP:conf/aaai/0001RV0BSW0G20}
\bibfield{author}{\bibinfo{person}{Sijia Liu}, \bibinfo{person}{Parikshit Ram},
  \bibinfo{person}{Deepak Vijaykeerthy}, \bibinfo{person}{Djallel Bouneffouf},
  \bibinfo{person}{Gregory Bramble}, \bibinfo{person}{Horst Samulowitz},
  \bibinfo{person}{Dakuo Wang}, \bibinfo{person}{Andrew Conn}, {and}
  \bibinfo{person}{Alexander~G. Gray}.} \bibinfo{year}{2020}\natexlab{}.
\newblock \showarticletitle{An {ADMM} Based Framework for AutoML Pipeline
  Configuration}. In \bibinfo{booktitle}{\emph{{AAAI}}}.
  \bibinfo{publisher}{{AAAI} Press}, \bibinfo{pages}{4892--4899}.
\newblock


\bibitem[Lu et~al\mbox{.}(2018)]%
        {DBLP:journals/pvldb/LuZSPZDHWPL18}
\bibfield{author}{\bibinfo{person}{Wei Lu}, \bibinfo{person}{Xinyi Zhang},
  \bibinfo{person}{Zhiyu Shui}, \bibinfo{person}{Zhe Peng},
  \bibinfo{person}{Xiao Zhang}, \bibinfo{person}{Xiaoyong Du},
  \bibinfo{person}{Hao Huang}, \bibinfo{person}{Xiaoyu Wang},
  \bibinfo{person}{Anqun Pan}, {and} \bibinfo{person}{Haixiang Li}.}
  \bibinfo{year}{2018}\natexlab{}.
\newblock \showarticletitle{{MSQL+:} a Plugin Toolkit for Similarity Search
  under Metric Spaces in Distributed Relational Database Systems}.
\newblock \bibinfo{journal}{\emph{Proc. {VLDB} Endow.}} \bibinfo{volume}{11},
  \bibinfo{number}{12} (\bibinfo{year}{2018}), \bibinfo{pages}{1970--1973}.
\newblock


\bibitem[Marcus and Papaemmanouil(2018)]%
        {DBLP:conf/sigmod/MarcusP18}
\bibfield{author}{\bibinfo{person}{Ryan Marcus} {and} \bibinfo{person}{Olga
  Papaemmanouil}.} \bibinfo{year}{2018}\natexlab{}.
\newblock \showarticletitle{Deep Reinforcement Learning for Join Order
  Enumeration}. In \bibinfo{booktitle}{\emph{aiDM@SIGMOD}}.
  \bibinfo{publisher}{{ACM}}, \bibinfo{pages}{3:1--3:4}.
\newblock


\bibitem[Pavlo et~al\mbox{.}(2019)]%
        {DBLP:journals/debu/PavloBJMMALS19}
\bibfield{author}{\bibinfo{person}{Andrew Pavlo}, \bibinfo{person}{Matthew
  Butrovich}, \bibinfo{person}{Ananya Joshi}, \bibinfo{person}{Lin Ma},
  \bibinfo{person}{Prashanth Menon}, \bibinfo{person}{Dana~Van Aken},
  \bibinfo{person}{Lisa Lee}, {and} \bibinfo{person}{Ruslan Salakhutdinov}.}
  \bibinfo{year}{2019}\natexlab{}.
\newblock \showarticletitle{External vs. Internal: An Essay on Machine Learning
  Agents for Autonomous Database Management Systems}.
\newblock \bibinfo{journal}{\emph{{IEEE} Data Eng. Bull.}}
  \bibinfo{volume}{42}, \bibinfo{number}{2} (\bibinfo{year}{2019}),
  \bibinfo{pages}{32--46}.
\newblock


\bibitem[Perera et~al\mbox{.}(2021)]%
        {DBLP:conf/icde/PereraORB21}
\bibfield{author}{\bibinfo{person}{R.~Malinga Perera}, \bibinfo{person}{Bastian
  Oetomo}, \bibinfo{person}{Benjamin I.~P. Rubinstein}, {and}
  \bibinfo{person}{Renata Borovica{-}Gajic}.} \bibinfo{year}{2021}\natexlab{}.
\newblock \showarticletitle{{DBA} bandits: Self-driving index tuning under
  ad-hoc, analytical workloads with safety guarantees}. In
  \bibinfo{booktitle}{\emph{{ICDE}}}. \bibinfo{publisher}{{IEEE}},
  \bibinfo{pages}{600--611}.
\newblock


\bibitem[Russo et~al\mbox{.}(2018)]%
        {DBLP:journals/ftml/RussoRKOW18}
\bibfield{author}{\bibinfo{person}{Daniel Russo}, \bibinfo{person}{Benjamin~Van
  Roy}, \bibinfo{person}{Abbas Kazerouni}, \bibinfo{person}{Ian Osband}, {and}
  \bibinfo{person}{Zheng Wen}.} \bibinfo{year}{2018}\natexlab{}.
\newblock \showarticletitle{A Tutorial on Thompson Sampling}.
\newblock \bibinfo{journal}{\emph{Found. Trends Mach. Learn.}}
  \bibinfo{volume}{11}, \bibinfo{number}{1} (\bibinfo{year}{2018}),
  \bibinfo{pages}{1--96}.
\newblock
\urldef\tempurl%
\url{https://doi.org/10.1561/2200000070}
\showDOI{\tempurl}


\bibitem[Sadri et~al\mbox{.}(2020)]%
        {DBLP:conf/icde/SadriGL20}
\bibfield{author}{\bibinfo{person}{Zahra Sadri}, \bibinfo{person}{Le
  Gruenwald}, {and} \bibinfo{person}{Eleazar Leal}.}
  \bibinfo{year}{2020}\natexlab{}.
\newblock \showarticletitle{Online Index Selection Using Deep Reinforcement
  Learning for a Cluster Database}. In \bibinfo{booktitle}{\emph{{ICDE}
  Workshops}}. \bibinfo{publisher}{{IEEE}}, \bibinfo{pages}{158--161}.
\newblock


\bibitem[Sharma and Dyreson(2022)]%
        {DBLP:conf/sac/SharmaD22}
\bibfield{author}{\bibinfo{person}{Vishal Sharma} {and}
  \bibinfo{person}{Curtis~E. Dyreson}.} \bibinfo{year}{2022}\natexlab{}.
\newblock \showarticletitle{Indexer++: workload-aware online index tuning with
  transformers and reinforcement learning}. In
  \bibinfo{booktitle}{\emph{{SAC}}}. \bibinfo{publisher}{{ACM}},
  \bibinfo{pages}{372--380}.
\newblock


\bibitem[Sharma et~al\mbox{.}(2021)]%
        {DBLP:conf/ideas/SharmaDF21}
\bibfield{author}{\bibinfo{person}{Vishal Sharma}, \bibinfo{person}{Curtis~E.
  Dyreson}, {and} \bibinfo{person}{Nicholas Flann}.}
  \bibinfo{year}{2021}\natexlab{}.
\newblock \showarticletitle{{MANTIS:} Multiple Type and Attribute Index
  Selection using Deep Reinforcement Learning}. In
  \bibinfo{booktitle}{\emph{{IDEAS}}}. \bibinfo{publisher}{{ACM}},
  \bibinfo{pages}{56--64}.
\newblock


\bibitem[Tan et~al\mbox{.}(2019)]%
        {DBLP:journals/pvldb/TanZLCZZQSCZ19}
\bibfield{author}{\bibinfo{person}{Jian Tan}, \bibinfo{person}{Tieying Zhang},
  \bibinfo{person}{Feifei Li}, \bibinfo{person}{Jie Chen},
  \bibinfo{person}{Qixing Zheng}, \bibinfo{person}{Ping Zhang},
  \bibinfo{person}{Honglin Qiao}, \bibinfo{person}{Yue Shi},
  \bibinfo{person}{Wei Cao}, {and} \bibinfo{person}{Rui Zhang}.}
  \bibinfo{year}{2019}\natexlab{}.
\newblock \showarticletitle{iBTune: Individualized Buffer Tuning for
  Large-scale Cloud Databases}.
\newblock \bibinfo{journal}{\emph{Proc. {VLDB} Endow.}} \bibinfo{volume}{12},
  \bibinfo{number}{10} (\bibinfo{year}{2019}), \bibinfo{pages}{1221--1234}.
\newblock


\bibitem[Wang et~al\mbox{.}(2022a)]%
        {DBLP:journals/chinaf/WangZCYZL22}
\bibfield{author}{\bibinfo{person}{Danshi Wang}, \bibinfo{person}{Chunyu
  Zhang}, \bibinfo{person}{Wenbin Chen}, \bibinfo{person}{Hui Yang},
  \bibinfo{person}{Min Zhang}, {and} \bibinfo{person}{Alan Pak~Tao Lau}.}
  \bibinfo{year}{2022}\natexlab{a}.
\newblock \showarticletitle{A review of machine learning-based failure
  management in optical networks}.
\newblock \bibinfo{journal}{\emph{Sci. China Inf. Sci.}} \bibinfo{volume}{65},
  \bibinfo{number}{11} (\bibinfo{year}{2022}).
\newblock


\bibitem[Wang et~al\mbox{.}(2021)]%
        {DBLP:journals/pvldb/WangTB21}
\bibfield{author}{\bibinfo{person}{Junxiong Wang}, \bibinfo{person}{Immanuel
  Trummer}, {and} \bibinfo{person}{Debabrota Basu}.}
  \bibinfo{year}{2021}\natexlab{}.
\newblock \showarticletitle{{UDO:} Universal Database Optimization using
  Reinforcement Learning}.
\newblock \bibinfo{journal}{\emph{Proc. {VLDB} Endow.}} \bibinfo{volume}{14},
  \bibinfo{number}{13} (\bibinfo{year}{2021}), \bibinfo{pages}{3402--3414}.
\newblock


\bibitem[Wang et~al\mbox{.}(2022b)]%
        {DBLP:conf/sigmod/WangZYDHDT0022}
\bibfield{author}{\bibinfo{person}{Zhaoguo Wang}, \bibinfo{person}{Zhou Zhou},
  \bibinfo{person}{Yicun Yang}, \bibinfo{person}{Haoran Ding},
  \bibinfo{person}{Gansen Hu}, \bibinfo{person}{Ding Ding},
  \bibinfo{person}{Chuzhe Tang}, \bibinfo{person}{Haibo Chen}, {and}
  \bibinfo{person}{Jinyang Li}.} \bibinfo{year}{2022}\natexlab{b}.
\newblock \showarticletitle{WeTune: Automatic Discovery and Verification of
  Query Rewrite Rules}. In \bibinfo{booktitle}{\emph{{SIGMOD} Conference}}.
  \bibinfo{publisher}{{ACM}}, \bibinfo{pages}{94--107}.
\newblock


\bibitem[Wang et~al\mbox{.}(2013)]%
        {DBLP:conf/ijcai/WangZHMF13}
\bibfield{author}{\bibinfo{person}{Ziyu Wang}, \bibinfo{person}{Masrour Zoghi},
  \bibinfo{person}{Frank Hutter}, \bibinfo{person}{David Matheson}, {and}
  \bibinfo{person}{Nando de Freitas}.} \bibinfo{year}{2013}\natexlab{}.
\newblock \showarticletitle{Bayesian Optimization in High Dimensions via Random
  Embeddings}. In \bibinfo{booktitle}{\emph{{IJCAI}}}.
  \bibinfo{publisher}{{IJCAI/AAAI}}, \bibinfo{pages}{1778--1784}.
\newblock


\bibitem[Warnell et~al\mbox{.}(2018)]%
        {DBLP:conf/aaai/WarnellWLS18}
\bibfield{author}{\bibinfo{person}{Garrett Warnell},
  \bibinfo{person}{Nicholas~R. Waytowich}, \bibinfo{person}{Vernon Lawhern},
  {and} \bibinfo{person}{Peter Stone}.} \bibinfo{year}{2018}\natexlab{}.
\newblock \showarticletitle{Deep {TAMER:} Interactive Agent Shaping in
  High-Dimensional State Spaces}. In \bibinfo{booktitle}{\emph{{AAAI}}}.
  \bibinfo{publisher}{{AAAI} Press}, \bibinfo{pages}{1545--1554}.
\newblock


\bibitem[Wu et~al\mbox{.}(2022a)]%
        {DBLP:journals/dase/WuLZZC22}
\bibfield{author}{\bibinfo{person}{Sai Wu}, \bibinfo{person}{Ying Li},
  \bibinfo{person}{Haoqi Zhu}, \bibinfo{person}{Junbo Zhao}, {and}
  \bibinfo{person}{Gang Chen}.} \bibinfo{year}{2022}\natexlab{a}.
\newblock \showarticletitle{Dynamic Index Construction with Deep Reinforcement
  Learning}.
\newblock \bibinfo{journal}{\emph{Data Sci. Eng.}} \bibinfo{volume}{7},
  \bibinfo{number}{2} (\bibinfo{year}{2022}), \bibinfo{pages}{87--101}.
\newblock


\bibitem[Wu et~al\mbox{.}(2022b)]%
        {DBLP:conf/sigmod/00010SWNCB22}
\bibfield{author}{\bibinfo{person}{Wentao Wu}, \bibinfo{person}{Chi Wang},
  \bibinfo{person}{Tarique Siddiqui}, \bibinfo{person}{Junxiong Wang},
  \bibinfo{person}{Vivek~R. Narasayya}, \bibinfo{person}{Surajit Chaudhuri},
  {and} \bibinfo{person}{Philip~A. Bernstein}.}
  \bibinfo{year}{2022}\natexlab{b}.
\newblock \showarticletitle{Budget-aware Index Tuning with Reinforcement
  Learning}. In \bibinfo{booktitle}{\emph{{SIGMOD} Conference}}.
  \bibinfo{publisher}{{ACM}}, \bibinfo{pages}{1528--1541}.
\newblock


\bibitem[Xia et~al\mbox{.}(2022)]%
        {DBLP:journals/cn/XiaZRW22}
\bibfield{author}{\bibinfo{person}{Qiufen Xia}, \bibinfo{person}{Lizhen Zhou},
  \bibinfo{person}{Wenhao Ren}, {and} \bibinfo{person}{Yi Wang}.}
  \bibinfo{year}{2022}\natexlab{}.
\newblock \showarticletitle{Proactive and intelligent evaluation of big data
  queries in edge clouds with materialized views}.
\newblock \bibinfo{journal}{\emph{Comput. Networks}}  \bibinfo{volume}{203}
  (\bibinfo{year}{2022}), \bibinfo{pages}{108664}.
\newblock


\bibitem[Yu et~al\mbox{.}(2020)]%
        {DBLP:conf/icde/Yu0C020}
\bibfield{author}{\bibinfo{person}{Xiang Yu}, \bibinfo{person}{Guoliang Li},
  \bibinfo{person}{Chengliang Chai}, {and} \bibinfo{person}{Nan Tang}.}
  \bibinfo{year}{2020}\natexlab{}.
\newblock \showarticletitle{Reinforcement Learning with Tree-LSTM for Join
  Order Selection}. In \bibinfo{booktitle}{\emph{{ICDE}}}.
  \bibinfo{publisher}{{IEEE}}, \bibinfo{pages}{1297--1308}.
\newblock


\bibitem[Yuan et~al\mbox{.}(2020)]%
        {DBLP:conf/icde/Yuan0FSH20}
\bibfield{author}{\bibinfo{person}{Haitao Yuan}, \bibinfo{person}{Guoliang Li},
  \bibinfo{person}{Ling Feng}, \bibinfo{person}{Ji Sun}, {and}
  \bibinfo{person}{Yue Han}.} \bibinfo{year}{2020}\natexlab{}.
\newblock \showarticletitle{Automatic View Generation with Deep Learning and
  Reinforcement Learning}. In \bibinfo{booktitle}{\emph{{ICDE}}}.
  \bibinfo{publisher}{{IEEE}}, \bibinfo{pages}{1501--1512}.
\newblock


\bibitem[Zhang et~al\mbox{.}(2019)]%
        {DBLP:conf/sigmod/ZhangLZLXCXWCLR19}
\bibfield{author}{\bibinfo{person}{Ji Zhang}, \bibinfo{person}{Yu Liu},
  \bibinfo{person}{Ke Zhou}, \bibinfo{person}{Guoliang Li},
  \bibinfo{person}{Zhili Xiao}, \bibinfo{person}{Bin Cheng},
  \bibinfo{person}{Jiashu Xing}, \bibinfo{person}{Yangtao Wang},
  \bibinfo{person}{Tianheng Cheng}, \bibinfo{person}{Li Liu},
  \bibinfo{person}{Minwei Ran}, {and} \bibinfo{person}{Zekang Li}.}
  \bibinfo{year}{2019}\natexlab{}.
\newblock \showarticletitle{An End-to-End Automatic Cloud Database Tuning
  System Using Deep Reinforcement Learning}. In
  \bibinfo{booktitle}{\emph{{SIGMOD} Conference}}. \bibinfo{publisher}{{ACM}},
  \bibinfo{pages}{415--432}.
\newblock


\bibitem[Zhang et~al\mbox{.}(2022a)]%
        {DBLP:journals/pvldb/ZhangCLWTLC22}
\bibfield{author}{\bibinfo{person}{Xinyi Zhang}, \bibinfo{person}{Zhuo Chang},
  \bibinfo{person}{Yang Li}, \bibinfo{person}{Hong Wu}, \bibinfo{person}{Jian
  Tan}, \bibinfo{person}{Feifei Li}, {and} \bibinfo{person}{Bin Cui}.}
  \bibinfo{year}{2022}\natexlab{a}.
\newblock \showarticletitle{Facilitating Database Tuning with Hyper-Parameter
  Optimization: {A} Comprehensive Experimental Evaluation}.
\newblock \bibinfo{journal}{\emph{Proc. {VLDB} Endow.}} \bibinfo{volume}{15},
  \bibinfo{number}{9} (\bibinfo{year}{2022}), \bibinfo{pages}{1808--1821}.
\newblock


\bibitem[Zhang et~al\mbox{.}(2021)]%
        {DBLP:conf/sigmod/ZhangWCJT0Z021}
\bibfield{author}{\bibinfo{person}{Xinyi Zhang}, \bibinfo{person}{Hong Wu},
  \bibinfo{person}{Zhuo Chang}, \bibinfo{person}{Shuowei Jin},
  \bibinfo{person}{Jian Tan}, \bibinfo{person}{Feifei Li},
  \bibinfo{person}{Tieying Zhang}, {and} \bibinfo{person}{Bin Cui}.}
  \bibinfo{year}{2021}\natexlab{}.
\newblock \showarticletitle{ResTune: Resource Oriented Tuning Boosted by
  Meta-Learning for Cloud Databases}. In \bibinfo{booktitle}{\emph{{SIGMOD}
  Conference}}. \bibinfo{publisher}{{ACM}}, \bibinfo{pages}{2102--2114}.
\newblock


\bibitem[Zhang et~al\mbox{.}(2022b)]%
        {DBLP:conf/sigmod/ZhangW0T0022}
\bibfield{author}{\bibinfo{person}{Xinyi Zhang}, \bibinfo{person}{Hong Wu},
  \bibinfo{person}{Yang Li}, \bibinfo{person}{Jian Tan},
  \bibinfo{person}{Feifei Li}, {and} \bibinfo{person}{Bin Cui}.}
  \bibinfo{year}{2022}\natexlab{b}.
\newblock \showarticletitle{Towards Dynamic and Safe Configuration Tuning for
  Cloud Databases}. In \bibinfo{booktitle}{\emph{{SIGMOD} Conference}}.
  \bibinfo{publisher}{{ACM}}, \bibinfo{pages}{631--645}.
\newblock


\bibitem[Zhou et~al\mbox{.}(2021)]%
        {DBLP:journals/pvldb/ZhouLCF21}
\bibfield{author}{\bibinfo{person}{Xuanhe Zhou}, \bibinfo{person}{Guoliang Li},
  \bibinfo{person}{Chengliang Chai}, {and} \bibinfo{person}{Jianhua Feng}.}
  \bibinfo{year}{2021}\natexlab{}.
\newblock \showarticletitle{A Learned Query Rewrite System using Monte Carlo
  Tree Search}.
\newblock \bibinfo{journal}{\emph{Proc. {VLDB} Endow.}} \bibinfo{volume}{15},
  \bibinfo{number}{1} (\bibinfo{year}{2021}), \bibinfo{pages}{46--58}.
\newblock


\bibitem[Zhou et~al\mbox{.}(2022)]%
        {DBLP:conf/icde/ZhouLLJLWF22}
\bibfield{author}{\bibinfo{person}{Xuanhe Zhou}, \bibinfo{person}{Luyang Liu},
  \bibinfo{person}{Wenbo Li}, \bibinfo{person}{Lianyuan Jin},
  \bibinfo{person}{Shifu Li}, \bibinfo{person}{Tianqing Wang}, {and}
  \bibinfo{person}{Jianhua Feng}.} \bibinfo{year}{2022}\natexlab{}.
\newblock \showarticletitle{AutoIndex: An Incremental Index Management System
  for Dynamic Workloads}. In \bibinfo{booktitle}{\emph{{ICDE}}}.
  \bibinfo{publisher}{{IEEE}}, \bibinfo{pages}{2196--2208}.
\newblock


\bibitem[Zou et~al\mbox{.}(2021)]%
        {DBLP:journals/pvldb/ZouDBIYJJ21}
\bibfield{author}{\bibinfo{person}{Jia Zou}, \bibinfo{person}{Amitabh Das},
  \bibinfo{person}{Pratik Barhate}, \bibinfo{person}{Arun Iyengar},
  \bibinfo{person}{Binhang Yuan}, \bibinfo{person}{Dimitrije Jankov}, {and}
  \bibinfo{person}{Chris Jermaine}.} \bibinfo{year}{2021}\natexlab{}.
\newblock \showarticletitle{Lachesis: Automated Partitioning for UDF-Centric
  Analytics}.
\newblock \bibinfo{journal}{\emph{Proc. {VLDB} Endow.}} \bibinfo{volume}{14},
  \bibinfo{number}{8} (\bibinfo{year}{2021}), \bibinfo{pages}{1262--1275}.
\newblock


\end{thebibliography}


\end{document}